\documentclass[11pt,a4paper]{article}
\pdfoutput=1
\usepackage{jcappub}

\usepackage{amsmath}
\usepackage{amssymb}
\usepackage{float}
\usepackage{wrapfig}
\usepackage{bm}
\usepackage{color}
\usepackage{rotating}
\usepackage{adjustbox}
\usepackage{comment}
\catcode`\@=11
\def\lsim{\mathrel{\mathpalette\@versim<}}
\def\gsim{\mathrel{\mathpalette\@versim>}}
\def\@versim#1#2{\vcenter{\offinterlineskip
\ialign{$\m@th#1\hfil##\hfil$\crcr#2\crcr\sim\crcr } }}
\catcode`\@=12

\usepackage{xcolor}
\usepackage{hyperref}
\usepackage{pdflscape}

\renewcommand{\figurename}{{\bf Figure}}
\renewcommand{\thefigure}{{\bf\arabic{figure}}}
\renewcommand{\tablename}{{\bf Table}}
\renewcommand{\thetable}{{\bf\arabic{table}}}

\hypersetup{
colorlinks=true,
citecolor=blue,
citebordercolor=red,
linktoc=all,
linkcolor=blue
}

\allowdisplaybreaks

\graphicspath{{./figs/}}

\begin{document}

\title{Universal gravitational-wave signatures from heavy new physics in the electroweak sector}

\author[a]{Astrid Eichhorn,}
\author[b]{Johannes Lumma,}
\author[b]{Jan M. Pawlowski,}
\author[a,c]{Manuel Reichert,}
\author[b]{Masatoshi Yamada}
\affiliation[a]{CP3-Origins, University of Southern Denmark, Campusvej 55, 5230 Odense M, Denmark}
\affiliation[b]{Institut f\"ur Theoretische Physik, Universit\"at Heidelberg, Philosophenweg 16, 69120 Heidelberg, Germany}
\affiliation[c]{Department  of  Physics  and  Astronomy,  University  of  Sussex,  Brighton,  BN1  9QH,  U.K.}
\emailAdd{eichhorn@cp3.sdu.dk}
\emailAdd{j.lumma@thphys.uni-heidelberg.de}
\emailAdd{j.pawlowski@thphys.uni-heidelberg.de}
\emailAdd{m.reichert@sussex.ac.uk}
\emailAdd{m.yamada@thphys.uni-heidelberg.de}

\abstract{
We calculate the gravitational-wave spectra produced by the electroweak phase transition with TeV-scale Beyond-Standard-Model physics in the early universe. Our study captures the effect of quantum and thermal fluctuations within a non-perturbative framework. We discover a universal relation between the mean bubble separation and the strength parameter of the phase transition, which holds for a wide range of new-physics contributions.

The ramifications of this result are three-fold: First, they constrain the gravitational-wave spectra resulting from heavy (TeV-scale) new physics. Second, they contribute to distinguishing heavy from light new physics directly from the gravitational-wave signature. Third, they suggest that a concerted effort of gravitational-wave observations together with collider experiments could be required to distinguish between different models of heavy new physics.
}

\maketitle 

\section{Introduction}
Gravitational-wave (GW) signals from a first-order phase transition \cite{Kosowsky:1991ua, Kosowsky:1992rz, Kosowsky:1992vn, Kamionkowski:1993fg, Allen:1996vm} constitute an intriguing link between particle physics and gravitational physics. If measured, these signals can be used to unveil the details of the evolution of the early universe during these phase transitions. Accordingly, the investigation of such GW- signals has attracted considerable attention, see \cite{Caprini:2015zlo, Mazumdar:2018dfl, Caprini:2018mtu, Caprini:2019egz} for reviews. 

The evolution of the early universe with a Standard Model (SM) high-energy sector features two phase transitions that could potentially trigger GW signals: The QCD phase transition has been proposed as a source of gravitational waves in \cite{Witten:1984rs}, but the transition is not of first-order, at least at low density. Similarly, the electroweak phase transition would only be of first order for a very low Higgs mass excluded by experiments, see \cite{Fodor:1994sj, Buchmuller:1995sf, Kajantie:1996mn, Rummukainen:1998as, Csikor:1998eu}. We conclude that a close-to-equilibrium evolution within the SM through these phase transitions does not trigger gravitational waves. 

Consequently, GW signals, that can be linked to a first-order phase transition, offer an exciting imprint of and evidence for Beyond-Standard-Model (BSM) physics. There are several potential sources for such a phase transition, most of them linked to yet unsolved high-energy-physics questions. For example, phase transitions in the dark sector could result in detectable GW signals, see, e.g., \cite{Grojean:2006bp, Schwaller:2015tja, Jaeckel:2016jlh, Kubo:2016kpb, Addazi:2017gpt, Tsumura:2017knk, Aoki:2017aws, Okada:2018xdh, Croon:2018erz, Breitbach:2018ddu, Huang:2020mso}.

In the present work, we concentrate on a potentially first-order electroweak phase transition triggered by BSM-contributions to the Higgs potential. It is well-known that new-physics models impact the Higgs potential and in particular, may change the order of the electroweak phase transition. Investigations have been done in supersymmetric models, \cite{Apreda:2001us}, and more recently, more generic new-physics models, e.g., models with one \cite{Profumo:2007wc,Espinosa:2011ax, Cline:2012hg, Profumo:2014opa, Vaskonen:2016yiu, Beniwal:2017eik, Beniwal:2018hyi, Hashino:2018wee} or several additional SM singlets \cite{Kakizaki:2015wua, Alves:2018oct, Alves:2018jsw, Alves:2019igs, Alves:2020bpi}, two-Higgs models \cite{Fromme:2006cm, Dorsch:2016nrg, Basler:2016obg} and composite Higgs models \cite{Konstandin:2011dr,Bruggisser:2018mrt, Miura:2018dsy}. At the LHC, new physics in the electroweak sector could be found by direct and indirect searches. Even if the mass scale of new physics is above the direct reach of the LHC, imprints of the new physics exist in the Higgs self-coupling \cite{Kanemura:2004ch, Noble:2007kk, Profumo:2014opa, Huang:2015izx, Huang:2015tdv, Kobakhidze:2015xlz, Hashino:2016rvx, Reichert:2017puo} that can be measured at a hadron collider~\cite{Baur:2002rb, Baur:2002qd, Butter:2016cvz}. 

Future GW interferometers like LISA and DECIGO offer an exciting complementary test of BSM physics. They will be sensitive to gravitational waves in the frequency range that is of interest for the electroweak phase transition \cite{Grojean:2006bp}; pulsar timing arrays are sensitive to lower frequencies, and could therefore find imprints of the QCD phase transition \cite{Caprini:2010xv}. In particular, a GW signature can provide information on the shape of the underlying potential~\cite{Chala:2019rfk}. 

By now, there is a plethora of works that explore the impact of BSM fields on the GW signature explicitly, e.g., \cite{Huber:2007vva, Konstandin:2011dr, Kakizaki:2015wua, Dorsch:2016nrg, Vaskonen:2016yiu, Chala:2016ykx, Artymowski:2016tme, Beniwal:2017eik, Chao:2017vrq, Bruggisser:2018mrt, Beniwal:2018hyi, Hashino:2018wee, Miura:2018dsy, Ellis:2019oqb, Dev:2019njv, Ellis:2020nnr}. Alternatively, the contributions of BSM-physics to the Higgs potential can be introduced in a phenomenological approach as mean-field contributions to the finite-temperature effective potential by hand, see, e.g., \cite{Wang:2020jrd} for a $\phi^6$ correction, as well as \cite{Ellis:2020awk} for a logarithmic correction. The zero-temperature one-loop contribution to the potential, arising from $\phi^6$ corrections has been investigated in \cite{Leitao:2015fmj}. Taking a step beyond, the one-loop quantum and thermal corrections have been explored for $\phi^6$ contributions to the microscopic potential \cite{Huber:2007vva, Delaunay:2007wb, Leitao:2015fmj, Cai:2017tmh, Ellis:2018mja, Ellis:2019oqb, Ellis:2020awk}. The uncertainties of standard daisy-resummed perturbation theory on the GW spectrum were discussed in~\cite{Croon:2020cgk}.

\begin{figure}[!t]
	\centering
	\includegraphics[width=.65\linewidth]{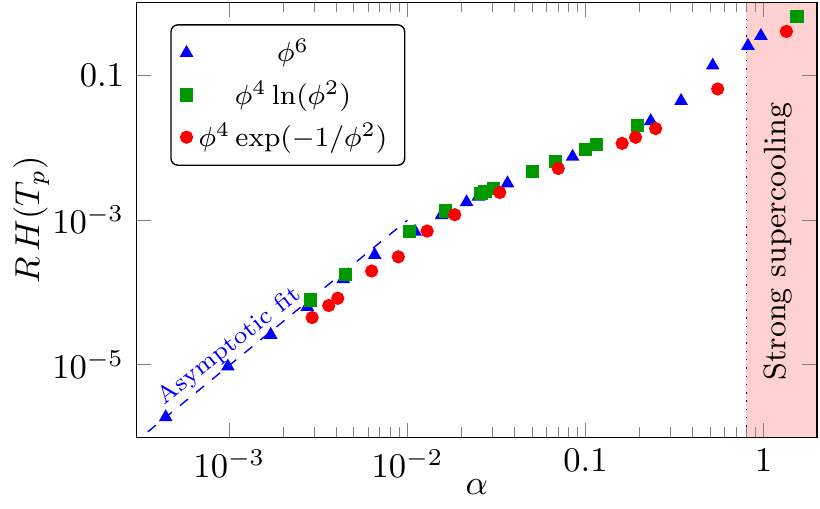}
	\caption{We display the strength parameter $\alpha$ versus the mean bubble separation $R$, multiplied by the Hubble parameter $H(T_p)$ at the percolation temperature $T_p$, as measured for the rather generic classes of new-physics models. We find an almost universal relation between $RH$ and $\alpha$ for all modifications. The red area marks the strong supercooling regime as defined in \autoref{sec:strong-supercooling}. At small $\alpha$, we find an asymptotic power-law behavior with $RH=10\, \alpha^{2}$.}
	\label{fig:HRAlphaPlane}
\end{figure}

In this paper, we provide a fully non-perturbative calculation of the parameters determining GW signals from a first-order electroweak phase transition. We consider rather generic new-physics models that arise from new physics in the TeV-scale range, see \autoref{fig:HRAlphaPlane}, and include their low-energy quantum and thermal fluctuations below the new-physics scale $M_{\rm NP}$ non-perturbatively. In these models, the Higgs potential at the scale $M_{\rm NP}$ is not well-approximated by a $\phi^4$-potential due to integrating out new massive particles with masses beyond $M_{\rm NP}$. To model the impact of such heavy new physics, we include several distinct classes of modification for the potential at $M_{\rm NP}$ that we expect to cover generic cases of new physics. Starting from these rather generic potentials at $M_{\rm NP}$, we integrated out quantum and thermal fluctuations below $M_{\rm NP}$ in \cite{Reichert:2017puo}, and use this quantum effective potential here to evaluate the GW signals from the electroweak phase transition in these models. Here, ``heavy" new physics refers to a new-physics scale higher than the Higgs mass, but low enough to have an impact on the electroweak phase transition. We work with $M_{\rm NP} = 2 \, \rm TeV$ throughout this paper, which corresponds to heavy new particles in comparison with those new-physics models where new degrees of freedom exist at or even below the Higgs mass.

The physics properties of the first-order transition, and hence that of the resulting GW signal, are characterized by two parameters, the strength parameter $\alpha$ (energy budget) and the mean bubble separation $R$, typically measured in units of the Hubble parameter $H(T_p)$ at the percolation temperature $T_p$. Importantly, we find that these two parameters are related by an almost universal curve over the whole parameter range. To highlight its importance this novel result is displayed already in \autoref{fig:HRAlphaPlane}, more details can be found in \autoref{sec:results}. As a consequence, the properties of the first-order phase transition which are important for the GW signal, are not sensitive to the specific form of the new-physics contribution at the ultraviolet scale $M_{\rm NP}$, but rather on 'integrated' information that is well described by either the transition strength $\alpha$ or the mean bubble separation. Quite surprisingly, this universality encompasses both non-perturbative as well as perturbative heavy new physics. This has ramifications both for GW searches as well as for a ``multi-messenger" approach which uses GW detectors and particle colliders concertedly.

This paper is structured as follows: In \autoref{sec:HOHiggs}, we introduce the theoretical framework we work in and recall the main results from \cite{Reichert:2017puo} for the electroweak phase transition that we build on here. In \autoref{sec:GWcalculation}, we review how to obtain the spectrum of gravitational waves from the finite-temperature effective Higgs potential. We present our results in \autoref{sec:results}, where we discover an intriguing universality between the strength parameter and the mean bubble separation, see \autoref{sec:universality}. Furthermore, we provide GW spectra for all modifications of the Higgs potential, see \autoref{sec:gw-spectra}, and we link the LISA signal-to-noise ratio with the high-luminosity LHC results on the Higgs self-couplings in \autoref{sec:GWandLHC}. We conclude in \autoref{sec:conclusions}.

\section{Higher-order contributions in the Higgs potential}
\label{sec:HOHiggs}
We build on the results of \cite{Reichert:2017puo}, where the impact of higher-order contributions in the Higgs potential to the Higgs self-couplings was studied and briefly review them here. We include modifications of the Higgs potential at a new-physics scale $ M_\text{NP}$. The new-physics scale is typically chosen in the TeV range, such that it impacts the physics of the electroweak phase transition. Lower new-physics scales are typically in tension with LHC results, whereas higher new-physics scales decouple the new physics from the electroweak phase transition. Starting from a given potential at $M_\text{NP}$, we successively integrate out quantum fluctuations to obtain the quantum effective potential of the Higgs field for a given finite temperature $T$. Following the evolution of the full quantum effective potential as a function of $T$ allows us to determine the order of the electroweak phase transition that results from a given new-physics modification at $M_\text{NP}$. We use the functional renormalization group (FRG) \cite{Wetterich:1992yh} as a non-perturbative tool, for reviews of the FRG, see, e.g., \cite{Berges:2000ew, Pawlowski:2005xe, Gies:2006wv, Delamotte:2007pf, Braun:2011pp, Dupuis:2020fhh}. In the context of Higgs physics, this method is particularly well suited to explore various questions related to the Higgs potential, such as its (non-perturbative) stability \cite{Gies:2013fua, Gies:2014xha, Eichhorn:2015kea, Borchardt:2016xju, Gies:2017zwf, Sondenheimer:2017jin}, as well as the impact of a portal to dark matter \cite{Eichhorn:2014qka, Held:2018cxd}. The method is well-suited to be used in the context of an effective theory with a finite new-physics scale, where it can in particular be used to go beyond perturbation theory and analyze generic new-physics contributions not restricted by perturbativity.

We do not work with a fully-fledged dynamical implementation of the whole SM, instead, we work in the framework of \cite{Eichhorn:2015kea}. We account for the effects of weak gauge bosons through a fiducial coupling and a thermal mass. The use of a fiducial coupling instead of a dynamical weak gauge sector has proven to reproduce the running of the SM up to the Planck scale \cite{Eichhorn:2015kea}. Accordingly, we also neglect the would-be Goldstone bosons and focus on a real scalar $\phi$ with $\mathbb{Z}_2$ reflection symmetry. In the spontaneously symmetry-broken phase, this scalar field relates directly to the physical Higgs field $\phi = H + v$, where $v$ denotes the vacuum expectation value (vev). In \cite{Reichert:2017puo}, the electroweak phase transition has been explored in this framework. 

\begin{figure}[!t]
 \includegraphics[width=0.99\linewidth]{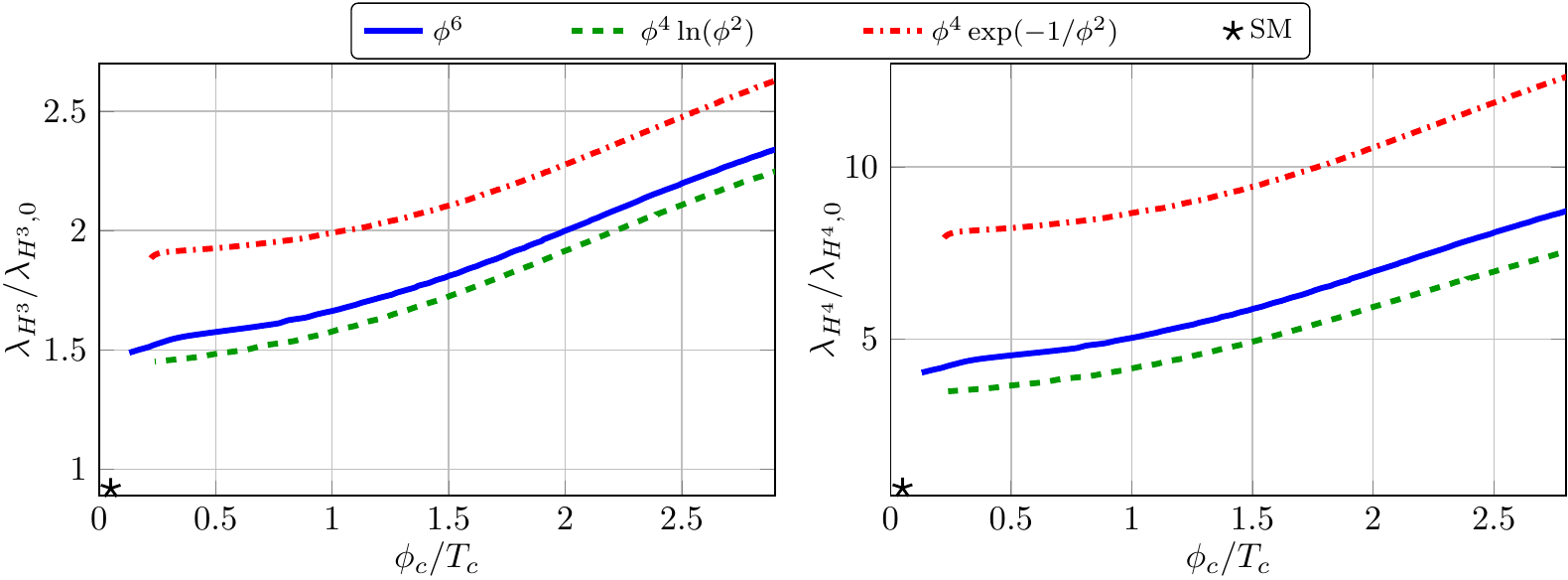}
 \caption{Three- and four-Higgs self coupling normalized to its tree-level value, $\lambda_{H^3}/\lambda_{H^3,0}$ and $\lambda_{H^4}/\lambda_{H^4,0}$, as a function of $\phi_c/T_c$ for all three modifications of the Higgs potential. The Higgs self-coupling increases with increasing strength of the phase transition. The results are taken from~\cite{Reichert:2017puo}.}
 \label{fig:Higgs-coupling}
\end{figure}

Heavy new physics beyond the mass threshold $M_\text{NP}$ can be parameterized by higher-order operators in the Higgs potential at and below $M_\text{NP}$. It is well known that the inclusion of such higher-order operators, e.g., a $\phi^6$ interaction, can lead to a strong first-order phase transition \cite{Grojean:2004xa, Bodeker:2004ws}. In \cite{Reichert:2017puo}, generic BSM contributions were investigated, all leading to a strong first-order phase transition. The BSM contributions, $\Delta V$, are added to the perturbatively renormalizable form of the potential at the new-physics scale, i.e.,
\begin{align}
	\label{eq:mod-Higgs-pot}
V_{k=M_\text{NP}} (\phi)= \frac{\mu^2}{2}\phi^2 + \frac{\lambda_4}{4}\phi^4 + \Delta V(\phi)\,.
\end{align}
The new physics contributions considered in \cite{Reichert:2017puo} were parameterized in three classes,
\begin{itemize}
\item Polynomial contributions: A $\Delta V(\phi) = \frac{\lambda_6}{24 M_\text{NP}^2}\phi^6$ interaction is the leading-order effect of new physics in a standard effective-field theory expansion. 
\item Logarithmic contributions: The modification to the effective action is logarithmic, as in the Coleman-Weinberg potential. This motivates modifications of the form $\Delta V(\phi) = \frac14 \lambda_{\ln} \phi^4\ln \frac{\phi^2}{2 M_\text{NP}^2}$.
\item Exponential contributions: Non-perturbative effects inspire an exponential dependence on $M_\text{NP}^2/\phi^2$, i.e., a contribution of the form $\Delta V(\phi) = \frac18 \lambda_{\exp} \phi^4 \exp\!\left(-\frac{2 M_\text{NP}^2}{\phi^2} \right)$.
\end{itemize}
These three classes rather generically cover different types of contributions that may arise from integrating out new physics above the new-physics scale $M_{\rm NP}$, which serves as an ultraviolet cutoff scale. They serve as initial potentials at the cutoff scale $M_\text{NP}$ \textit{before} integrating out quantum fluctuations below $M_{\rm NP}$. We numerically evaluate the running of the scalar potential $V_k(\phi)$ at finite temperature $T$ on a grid in the scalar field $\phi$, for more details see \cite{Reichert:2017puo}. Let us highlight that in \cite{Reichert:2017puo}, it was confirmed explicitly that -- as expected -- a $\phi^8$ contribution is well captured by a $\phi^6$-modification, and similarly, a $\phi^4\ln \phi^2$ contribution resembles a $\phi^2\ln \phi^2$ one, and finally, a $\phi^6 \exp(-\phi^{-2})$ contribution can be encoded in a $\phi^4 \exp(-\phi^{-2})$ one. Therefore, we restrict ourselves to the three generic choices listed above for the present work and assume that these provide us with representative examples for entire classes of new-physics modifications. We stress that the form of $\Delta V(\phi)$ as listed above holds at the new-physics scale $M_{\rm NP}$. Towards lower scales, the RG flow automatically generates additional higher-order contributions that are included in our potential at lower scales, see \autoref{App:FRG} for more details.

All three types of modifications can change the electroweak phase transition from a cross over in the SM to a first-order phase transition. The strength of the phase transition is parameterized by the ratio $\phi_c/T_c$, with $\phi_c = \langle \phi\rangle_{T_c}$ the expectation value of the quantum field at the nontrivial minimum at the critical temperature $T_c$. A strong first-order phase transition with $\phi_c/T_c>1$ can be achieved for all three types of modifications. This requires that the new physics contributes significantly to the Higgs potential at temperatures around $T_c$. This in turn implies that the new-physics scale $M_\text{NP}$ cannot be too far above the TeV range, since all three types of modifications $\Delta V$ are suppressed under the RG flow to the IR due to their higher-order nature. Despite the non-perturbative nature of the second and third class of modifications, the logarithmic and exponential ones, this perturbative suppression of higher-order terms holds for the set of parameters we explore. In \cite{Reichert:2017puo}, it was estimated that $ M_\text{NP}\approx 10$\,TeV is the maximal new-physics scale that can still impact the electroweak phase transition at fixed Higgs mass and vev. 

In \cite{Reichert:2017puo}, the modifications were connected to two LHC observables, the effective three-Higgs coupling $\lambda_{H^3}$ and four-Higgs coupling $\lambda_{H^4}$. A similar connection has been made in the literature in \cite{Kanemura:2004ch, Noble:2007kk, Huang:2015izx, Huang:2015tdv, Kobakhidze:2015xlz, Hashino:2016rvx}. Due to the modifications of the scalar potential at $M_\text{NP}$, the three- and four-Higgs couplings are enhanced, compared to their values in the SM without such modifications. We display this in \autoref{fig:Higgs-coupling}. With the high-luminosity LHC run, a modification of more than $\lambda_{H^3}/\lambda_{H^3,0}>1.7$ will be detectable at 68\% confidence level \cite{Kling:2016lay}. Therefore, the LHC may provide \emph{indirect} evidence for a first-order phase transition. One of our main aims in this work is to connect this LHC observable with GW signatures of a first-order electroweak phase transition that will become detectable at future, space-based GW detectors, such as the planned observatory LISA and the proposed observatory DECIGO. Similar links between GW detectors and collider searches have been established within other new-physics settings, e.g., in \cite{Artymowski:2016tme, Beniwal:2017eik, Chala:2018ari, Hashino:2018wee, Zhou:2020idp}.

As mentioned before, the above three classes span, rather generically, a wide range of possible modifications to the potential. Exploring these three classes allows us to map out key parts of the parameter space relevant for GWs. Ultimately, this provides us with an estimate of achievable signal-to-noise-ratio, peak frequency, and peak amplitude for GW signals from an electroweak first-order phase transition driven by heavy new physics.

\section{Review: Calculation of gravitational-wave spectra from 1st-order phase transition}
\label{sec:GWcalculation}
In this section, we review the calculation of the GW spectrum from a first-order phase transition and explain the assumptions and approximations we make to obtain GW spectra. We aim at providing a self-contained discussion of this topic; reviews can be found in~\cite{Cai:2017cbj, Weir:2017wfa, Caprini:2018mtu, Caprini:2019egz, Wang:2020jrd, Hindmarsh:2020hop}. At the end of this section, we will summarize which equations we are using for the computation of the GW spectra.

During a first-order phase transition, bubbles of the true vacuum are produced in the false vacuum. These bubbles expand and eventually collide with each other~\cite{Kosowsky:1991ua, Kosowsky:1992rz, Kosowsky:1992vn, Kamionkowski:1993fg, Caprini:2007xq, Huber:2008hg, Caprini:2009fx, Espinosa:2010hh, Weir:2016tov, Jinno:2016vai}. After the initial collision, sound shells continue to propagate in the plasma~\cite{Hindmarsh:2013xza, Giblin:2013kea, Giblin:2014qia, Hindmarsh:2015qta, Hindmarsh:2017gnf}. Moreover, a first-order transition can generate magneto-hydrodynamical turbulence in the cosmic plasma~\cite{Kosowsky:2001xp,Dolgov:2002ra, Caprini:2006jb, Gogoberidze:2007an, Kahniashvili:2008pe, Kahniashvili:2009mf, Caprini:2009yp, Kisslinger:2015hua}. All three processes trigger tensor fluctuations in the energy-momentum tensor that describes the primordial plasma, and therefore they source GWs. The process that typically adds the largest contribution to the GW signal is the propagation of sound waves in the plasma \cite{Hindmarsh:2013xza, Hindmarsh:2015qta}.

For all three components, fits for the resulting spectrum of GWs are available, see, e.g., \cite{Weir:2017wfa}. The most important information is the peak frequency $f_\text{peak}$ and the peak amplitude $h^2 \Omega^\text{peak}$. The peak frequency depends on the inverse duration time of the phase transition, as expected on dimensional grounds. The amplitude of the GW signal depends on the amount of energy that is released. Both quantities can be calculated from the finite-temperature effective potential.

\subsection{Characteristic temperatures}
To calculate GW spectra, several different temperatures that characterize the phase transition in an expanding universe are relevant:
\begin{itemize}
\item $T_c$ is the critical temperature, at which both minima of the effective potential are degenerate, i.e.,
\begin{align}
 V_\text{eff}(\phi=0,T_c) = V_\text{eff}(\phi=\phi_c,T_c),
\end{align}
where $\phi_c=\langle \phi \rangle_{T_c}$ is the vacuum expectation value in the broken phase at $T_c$.
\item $T_n$ is the nucleation temperature, which is determined by comparing the decay rate $\Gamma(t)$ of the false vacuum to the expansion rate of the universe, described by the Hubble parameter $H(t)$. At nucleation temperature, one bubble nucleation per causal Hubble volume takes place on average, i.e.,
\begin{align}
 N(T_n)= \int_{t_c}^{t_n} \! \mathrm dt \frac{\Gamma(t)}{H(t)^3}= 1\,.
 \label{eq:nucleationtemp}
\end{align}
Here, $t_c$ is the time at which the temperature equals the critical temperature. This is the time at which bubble nucleation can in principle start.
\item The percolation temperature $T_p$ is defined as the temperature at which the probability to have the false vacuum is about 0.7; i.e., 34 \% of the false vacuum has been converted into the true vacuum.
\end{itemize}
Additionally, two further temperatures can play a role, namely the reheating temperature and the minimization temperature, both of which will be defined later.

The order of the temperatures  is always $T_p \leq T_n \leq T_c$. In a cosmological context, the phase transition does not occur at $T_c$, but the cosmological phase transition temperature is the percolation temperature $T_p$. Supercooling can occur when the expansion rate of the universe is large enough to strongly suppress the tunneling probability, even if the true vacuum lies at a significantly lower value of the potential than the false vacuum. Strong supercooling can therefore enhance the strength of the phase transition and the amount of energy that is released. When a phase transition is strongly supercooled, it becomes important to accurately define the criterion for the phase transition and to distinguish the characteristic temperatures, which is explained in the following sections. Last, if the expansion rate of the universe becomes too large, the criteria for the nucleation and percolation temperature might never be fulfilled and the phase transition might not complete. 

\subsection{Bubble nucleation and Euclidean action}
In first-order phase transitions, expanding bubbles of the broken phase are stochastically created through tunneling from the false vacuum. The nucleation rate for bubbles of the true vacuum, or equivalently the decay rate of the false vacuum, takes the following form as a function of time $t$~\cite{Coleman:1977py}:
\begin{align}
 \Gamma(t) = A(t)\,e^{-S(t)}.
 \label{nucleation rate of a bubble}
\end{align}
At much lower temperatures than the typical energy scale of the system, $S(t)$ is given by the four-dimensional Euclidean action, denoted here by $S_4(t)$, and the pre-factor is $A(t)= r_0^{-4}[S_4(t)/(2\pi)]^2$ where $r_0$ is the initial radius of the nucleated bubble. At finite temperature, the bubbles of true vacuum are induced by thermal fluctuations, which are dominating over the quantum fluctuations. We have checked that this also holds in our computation and from here on we focus on the thermal fluctuations. The nucleation rate \eqref{nucleation rate of a bubble} is determined by $A(T)=T^4[S_3(T)/(2\pi T)]^{3/2}$ and $S(T)=S_3(T)/T$~\cite{Linde:1980tt,Linde:1981zj}. Here $S_3$ is the three-dimensional Euclidean action, which reads
\begin{align}
S_3(T) =\int \! \mathrm d^3x \left[ \frac{1}{2}(\nabla \phi)^2 + V_\text{eff}(\phi,T)\right]
=4\pi \int^\infty_0 \! \mathrm dr\,r^2\left[ \frac{1}{2}\left( \frac{\mathrm d\phi}{\mathrm dr}\right)^2 +V_\text{eff}(\phi,T) \right],
\label{three dimensional Euclidean action}
\end{align}
where $r=\sqrt{x^2+y^2+z^2}$ is the three-dimensional distance. To evaluate $S_3(T)$, we first need to evaluate $\phi(r)$ by solving
\begin{align}
\frac{\mathrm d^2\phi(r)}{\mathrm dr^2} +\frac{2}{r}\frac{\mathrm d\phi(r)}{\mathrm dr} =\frac{\partial V_\text{eff}}{\partial \phi}\,,
\label{bounce equation}
\end{align}
with the boundary conditions, $\phi(r\to \infty) = 0$ and $\mathrm d\phi(r=0)/\mathrm dr=0$. The solution to \eqref{bounce equation} is the so-called bounce solution, which provides the radius of a bubble at temperature $T$. One can solve \eqref{bounce equation} using the overshooting/undershooting method. In this work, we employ the Python code {\sf CosmoTransitions}~\cite{Wainwright:2011kj} to find the bounce solution. 

The expansion history of the universe can be mapped onto its thermal history through the relation
\begin{align}
\frac{\mathrm d T}{\mathrm d t}=-H(T) T,
\label{adiabatic time temperature relation}
\end{align}
where $H(T)$ is the Hubble parameter, given by
\begin{align}
H(T)=\sqrt{\frac{\rho_\text{rad}+\rho_\text{vac}}{3M_\text{pl}^2}},
\label{eq:Hubble}
\end{align}
with the reduced Planck mass squared, $M_\text{pl}=2.435 \cdot 10^{18}$\,GeV. The radiation and vacuum energy densities are given by
\begin{align}
\rho_\text{rad}&= g_*\frac{\pi^2}{30} T^4\,, 
&
\rho_\text{vac}&= \Delta V_\text{eff}(\phi,T) = -[V_\text{eff}(\phi=\langle \phi \rangle_T,T) - V_\text{eff}(\phi=0,T)]\,,
\label{energy density of vacuum}
\end{align}
where $\langle \phi \rangle_T$ denotes the true vacuum at temperature $T$. The factor $g_*$ denotes the effective number of relativistic degrees of freedom, given by
\begin{align}
g_* = \sum_{i=\text{boson}}g_i \left(\frac{T_i}{T}\right)^4 + \frac{7}{8} \sum_{i=\text{fermion}}g_i \left(\frac{T_i}{T}\right)^4,
\end{align}
where $g_i$ and $T_i$ are the number of degrees of freedom and the decoupling temperature of particle species $i$, respectively. We treat the number of relativistic degrees of freedom as a constant value, $g_*=106.75$ in the SM, although in general it depends on the temperature. The new degrees of freedom that generate the modification of the Higgs potential at $M_\text{NP}$ are heavy and thus do not contribute to $g_*$ at the temperature of the electroweak phase transition. Using \eqref{adiabatic time temperature relation}, we can rewrite dependences on time in terms of temperature, so that hereafter we freely exchange between dependences on time and temperature. In particular, we can use it to rewrite \eqref{eq:nucleationtemp}, the defining equation for the nucleation temperature, i.e., the number of bubbles per causal Hubble volume at nucleation temperature, $N(T_n)$ in terms of  the nucleation temperature itself:
\begin{align}
N(T_n)=\int_{t_c}^{t_n}\mathrm dt \frac{\Gamma(t)}{H(t)^3}=\int_{T_n}^{T_c}\frac{\mathrm dT}{T}\frac{\Gamma(T)}{H(T)^4}=1\,.
\label{rate between nucleation and Hubble}
\end{align}
For fast phase transitions the integral in \eqref{rate between nucleation and Hubble} is dominated by $T\approx T_n$ and we arrive at the approximation
\begin{align}
\frac{\Gamma(T_n)}{H(T_n)^4}\approx 1 \,, 
\label{criterion for nucleation temperature}
\end{align}
which is commonly used.
Using the definitions of $\Gamma(T)$ and $H(T)$ given above, the relation \eqref{criterion for nucleation temperature} approximately yields 
\begin{align}
\frac{S_3(T_n)}{T_n}  \simeq 4\log (M_\text{pl}/T_n) \,.
\label{criterion for nucleation temperature in terms of action}
\end{align}
Assuming that the nucleation temperature is located in the range $10^2\,\text{GeV} \lsim T_n \lsim 10^4\,\text{GeV}$, one has $140\lsim S_3(T_n)/T_n \lsim 150$. These approximations do not hold for phase transitions with strong supercooling. In these cases, we resort to the definition \eqref{rate between nucleation and Hubble} of the nucleation temperature.

The time evolution of the universe is crucial to determine the percolation temperature. In particular, the phase transition occurs at roughly $T_n$ in the case of a fast phase transition, whereas, in general cases,  one should give a more accurate prescription. To that end, we define the false vacuum probability, in order to describe the percolation of bubbles~\cite{Guth:1979bh,Guth:1981uk},
\begin{align}
P(T)= e^{-I(T)} \,.\label{eq:falsevacprob}
\end{align}
Here, the weight function is given by~\cite{Ellis:2018mja}
\begin{align}
I(t)&=\int^t_{t_c} \mathrm dt'\,\Gamma(t') \frac{a(t')^3}{a(t)^3}  \left[ \frac{4\pi}{3}r(t',t)^3\right],
\label{weight of the false vacuum in terms of time}
\end{align}
where $a(t)$ is the scale factor in the Friedmann-Robertson-Walker metric and $r(t',t)$ is the comoving radius of a bubble that is nucleated at $t''(T'')$ and evolutes until $t(T)$. In terms of the wall velocity $v_w(T)$, the comoving bubble radius is given by
\begin{align}
r(t',t)= \int^t_{t'} \!\mathrm dt''\,v_w(t'') \frac{a(t)}{a(t'')}\,.
\label{comoving radius of a bubble}
\end{align}
Note that the scale factors in \eqref{weight of the false vacuum in terms of time} play a role in the dilution of nucleated bubbles. Using \eqref{adiabatic time temperature relation}, the weight function \eqref{weight of the false vacuum in terms of time} can be rewritten in terms of temperature
\begin{align}
I(T)=\frac{4\pi}{3} \int^{T_c}_T \! \mathrm dT'\frac{\Gamma(T')}{H(T')T'{}^4} \left( \int^{T'}_{T}\mathrm dT''\frac{v_w(T'')}{H(T'')}\right)^3 .
\label{weight of the false vacuum}
\end{align}
The percolation of bubbles starts when $I(T)\gsim0.34$ (corresponding to $P(T)\lsim 0.7$)~\cite{Rintoul_1997}. Hence, the percolation temperature $T_p$ is defined such that $I(T_p)\simeq 0.34$.

\subsection{Mean bubble separation}
The mean bubble separation $R(t)$ is a relevant length scale for the generation of gravitational waves produced by the phase transition. The number of bubbles per horizon is estimated from the averaged nucleation rate $\bar \Gamma(t)$ obtained from the false vacuum probability $P(T)$, cf.~\eqref{eq:falsevacprob}, and the decay rate of the false vacuum $\Gamma(t)$, cf.~\eqref{nucleation rate of a bubble},
\begin{align}
\bar \Gamma(t)=P(t)\Gamma(t) \,.
\label{averaged nucleation rate}
\end{align}
As the universe gradually fills up with bubbles of the true vacuum, the bubble nucleation is suppressed. Thus the number density of bubbles at time $t$ is given by~\cite{Enqvist:1991xw,Turner:1992tz},
\begin{align}
n_\text{B}(t)= \int^t_{t_c} \mathrm dt' \frac{a(t')^3}{a(t)^3} \bar \Gamma(t')\,,
\label{the number density of bubbles}
\end{align}
from which the mean bubble separation $R(t)$ at $t$ is defined as
\begin{align}
R(t) =(n_\text{B}(t))^{-1/3}\,.
\label{Mean bubble separation from the number density of bubbles}
\end{align}
It corresponds to the average distance between centers of nucleation points. We estimate $R(t)$ for specific cases below.

\subsection{Energy released by the first-order phase transition}
The amplitude of the GW signal strongly depends on the energy budget of the phase transition, which is commonly described by the strength parameter $\alpha$. Most analyses employ the bag model, where the bag constant $\epsilon$ describes the jump in energy and pressure across the phase boundary. The strength parameter is then given by $\alpha = \epsilon/ (a_+ T^4)$ evaluated at the percolation temperature, where $a_+= \pi^2 g_\text{eff}/30$ relates to the relativistic degrees of freedom in the symmetric phase. The remaining question is how to relate a concrete particle-physics model to the bag model. The currently most typically employed link uses that the bag constant $\epsilon$ is related to the trace of the energy-momentum tensor $\theta = (e-3p)/4$, where the energy density is given by the (0,0) component of the stress-energy tensor, $e=T_{00}$, and the pressure corresponds to the spatial entries, $p= T_{ii}$, with $i=1,2,3$ and no summation implied, as usual for a perfect fluid. Thus the definition of $\alpha$ is
\begin{align}
\alpha_\theta = \frac 34 \frac{\theta_+ - \theta_-}{a_+ T^4}\bigg|_{T=T_p},
\label{eq:alpha-theta}
\end{align}
where $+$ labels the symmetric vacuum and $-$ the symmetry-broken vacuum. Within the field-theoretic description, the energy density and pressure can be obtained from the effective potential, $e_{+/-}= V_\text{eff} - T \,\partial V_\text{eff}/\partial T$ and $p_{+/-}=-V_\text{eff}$. This leads us to
\begin{align}
\alpha_\theta 
= \frac{30}{\pi^2 g_\text{eff} T^4} \left(\Delta V_\text{eff} - \frac{T}{4} \frac{\partial \Delta V_\text{eff}}{\partial T} \right)\bigg|_{T=T_p}
=\frac{1}{\rho_\text{rad}} \left(\rho_\text{vac} - \frac{T}{4} \frac{\partial \rho_\text{vac}}{\partial T} \right)\bigg|_{T=T_p}.
\label{basic quantity: alpha: free energy}
\end{align}
Here, $\rho_\text{vac}$ and $\rho_\text{rad}$ are defined in \eqref{energy density of vacuum}. The parameter $\alpha_\theta$ is the ratio between the latent heat and the radiation energy and measures the strength of the phase transition in a way that is directly relevant for the GW production. 

For a weakly supercooled phase transition, i.e., $T_p\lsim T_c$, the vacuum-energy density $\rho_\text{vac}$ can be ignored, while for a strongly supercooled case, $T_p \ll T_c$, the vacuum-energy density becomes much larger than the radiation-energy density, such that \eqref{basic quantity: alpha: free energy} can be approximated by
\begin{align}
\alpha_\theta \simeq \frac{\rho_\text{vac}}{\rho_\text{rad}}\bigg|_{T=T_p} \,.
\end{align}

An alternative way to match the strength parameter to the bag model is via the energy difference of the two phases
\begin{align}
\alpha_e = \frac 34 \frac{e_+ - e_-} {a_+ T^4}\bigg|_{T=T_p} 
=  \frac{1}{\rho_\text{rad}} \left(\rho_\text{vac} - T \frac{\partial \Delta V_\text{eff}}{\partial T} \right)\bigg|_{T=T_p} \,,
\label{traditional definition of alpha}
\end{align}
which differs by a factor of four in the second term compared to \eqref{basic quantity: alpha: free energy}. The difference is due to the fact that the pressure-contribution to $\theta$, which is present in \eqref{eq:alpha-theta}, is ignored in the above definition based on the energy densities alone. This change of definition can be partially compensated, if we properly adjust  the corresponding efficiency parameter, which we define in \autoref{sec:GW-spectra}. Nonetheless, $\alpha_\theta$ and $\alpha_e$ are similar for strong phase transitions, i.e., where $\alpha$ is large and the vacuum energy dominates, while $\alpha_e$ overestimates the strength of weak phase transitions by a factor of four. For a discussion of the strength parameter beyond the bag model, see~\cite{Giese:2020rtr}.

\subsection{First-order phase transitions without strong supercooling}
For sufficiently fast phase transitions, the bubble nucleation rate at finite temperature can be approximated as
\begin{align}
\Gamma \approx \Gamma(t_0)  e^{\beta (t-t_0)\,+\,\dots} \,,
\label{eq:LinearBeta}
\end{align}
for some $t_0$, typically taken as the percolation time $t_p$. Herein $\beta$ can be understood as the inverse duration of the phase transition. The nucleation rate grows exponentially. From a Taylor expansion of \eqref{nucleation rate of a bubble} with $S = S_3(T)/T$, it follows that the timescale is set by
\begin{align}
\beta= -\frac{\mathrm d}{\mathrm dt}\frac{S_3(T)}{T}\bigg|_{t=t_0} \,.
\label{eq:LinearBeta2}
\end{align}
Equivalently, using \eqref{adiabatic time temperature relation}, the inverse duration of the phase transition can be written as
\begin{align}
\tilde\beta:= \frac{\beta}{H}\bigg|_{T=T_0} = T \frac{\mathrm dS}{\mathrm dT}\bigg|_{T=T_0} \,,
\label{eq:LinearBeta3}
\end{align}
where the expression is evaluated at the temperature corresponding to $t_0$. Note that \eqref{eq:LinearBeta2} and \eqref{eq:LinearBeta3} only hold if the duration of the phase transition is short enough such that the linearization of the action is a valid approximation. 

In fast phase transitions, for which the bubble-wall velocity and the scale factor are approximately constant, the comoving radius of a bubble \eqref{comoving radius of a bubble} is given by $r(t,t')=v_w \cdot(t-t')$. Then, with \eqref{eq:LinearBeta}, we obtain the weight function $I(t)$ in \eqref{weight of the false vacuum in terms of time} and the averaged nucleation rate $\bar\Gamma(t)$ in \eqref{averaged nucleation rate} in terms of $\beta$,
\begin{align}
I(t)&=I_0 e^{\beta (t-t_0)} \,,
&
\bar\Gamma(t)&= \Gamma_0 e^{\beta (t-t_0)}e^{-I_0 e^{\beta(t-t_0)}}\,.
\end{align}
Here, we use that $t_c \ll t_p$ and thus we can take the limit $t_c \rightarrow -\infty$ as a simplification. In the above expressions, we have defined $\Gamma_0=\Gamma(t_0)$ and $I_0=8\pi v_w^3\Gamma_0\beta^{-4}$. Thus, the number density \eqref{the number density of bubbles} can be expressed in terms of $\beta$ as
\begin{align}
n_\text{B}(t)= \frac{\beta^3}{8\pi v_w^3}\left[ 1- P(t)\right].
\end{align}
For $t>t_p$, the phase transition completes very fast and hence $P(t>t_p)\simeq 0$. Thus the mean bubble separation \eqref{Mean bubble separation from the number density of bubbles} at $t=t_p$ ($T= T_p$) is given by~\cite{Enqvist:1991xw}
\begin{align}
 R = (8\pi)^{\frac 13}\,v_w\beta^{-1}\,,
\label{mean bubble separation in fist phase transition}
\end{align}
To calculate the GW spectra for phase transitions without strong supercooling, the inverse duration $\beta$ and the mean bubble separation $R$ can therefore be used interchangeably.

\subsection{Strongly supercooled first-order phase transitions}
\label{sec:strong-supercooling}
For sufficiently strong phase transitions, $\alpha \sim \mathcal{O}(1)$, the weight $S_3(T)/T$ has a minimum at $T_m$ and increases for $T<T_m$. This defines the minimization temperature $T_m$, which is only important in the situation of strong supercooling.

In this case, $\beta$ as defined above goes to zero at the minimization temperature $T_m$ and can even become negative for $T<T_m$, such that the approximation in \eqref{eq:LinearBeta} breaks down~\cite{Megevand:2016lpr,Jinno:2017ixd,Cutting:2018tjt}. For a more robust determination of the duration of the phase transition, one has to consider the quadratic term in $(t-t_m)$ in the Taylor expansion of the nucleation rate in \eqref{eq:LinearBeta}, i.e.,
\begin{align}
\Gamma \propto e^{-\frac{S_3}{T}}= e^{\frac{1}{2}\beta_V^2 (t-t_m)^2\,+\dots}\,,
\label{next order approximation in the nucleation rate}
\end{align}
where we used that for sufficiently strong phase transitions, $\beta\rightarrow 0$. The duration of the phase transition is now set by the second derivative of $S_3/T$ with respect to $t$, i.e.,
\begin{align}
\beta_V\equiv \left.\sqrt{-\frac{\mathrm d^2}{\mathrm dt^2}\left(\frac{S_3(T)}{T}\right)}\right|_{t=t_m}
=\left. H(T)T\sqrt{\frac{\mathrm d^2}{\mathrm dT^2}\left(\frac{S_3(T)}{T}\right)}\right|_{T=T_m}\,.
\label{eq:beta-strong-supercooling}
\end{align}
In the next sections, we will use a common parameter $\tilde{\beta}$ to denote $\beta/H$ in cases without strong supercooling, and $\beta_V/H$ in cases with strong supercooling.

Next, we evaluate the mean bubble separation in the case of a strongly supercooled first-order phase transition.
For the initial nucleation rate \eqref{next order approximation in the nucleation rate}, when the false vacuum probability satisfies $P \simeq 1$, the number density is given by~\cite{Megevand:2016lpr}
\begin{align}
n_\text{B}(t)=n_\text{max}\frac{1+\text{erf}\!\left[\beta_V(t-t_m)/\sqrt{2} \right]}{2}\exp\!\left[ -3H(t-t_m) +\left(\frac{3}{\sqrt{2}}\frac{H}{\beta_V} \right)^2 \right],
\label{number density in the strongly supercooled first-order phase transition cases}
\end{align}
where $n_\text{max}=\sqrt{2\pi}\Gamma(t_m)\beta_V^{-1}$ and erf denotes the error function. In the derivation of \eqref{number density in the strongly supercooled first-order phase transition cases}, a constant Hubble parameter is assumed. This approximation holds in the case of a strongly supercooled first-order phase transition for which the vacuum energy dominates, see \eqref{eq:Hubble} and \eqref{energy density of vacuum}, where $\rho_\text{rad}$ is negligible and $\rho_\text{vac}$ is roughly temperature independent. For $t-t_m\sim \sqrt{2}\beta_V^{-1}$, we obtain the mean bubble separation~\cite{Cutting:2018tjt}
\begin{align}
R = (n_\text{max})^{-1/3}= \left[\sqrt{2\pi}\, \Gamma(T_m)\beta_V^{-1} \right]^{-\frac{1}{3}}\,.
\label{eq:meanbubsepstrongssc}
\end{align}
The relation between mean bubble separation and the inverse duration of the phase transition contains the decay rate evaluated at the minimization temperature, requiring knowledge about the effective scalar potential. In comparison, in the case without strong supercooling, the relation does not require this piece of information, see \eqref{mean bubble separation in fist phase transition}.

\subsection{Maximal strength of the phase transition}
\label{sec:max-strength}
The requirement that the phase transition needs to be completed within an expanding universe results in a maximum possible strength of the electroweak phase transition. A na\"ive criterion for the completion of the phase transition is that the percolation of bubbles takes place as defined below \eqref{weight of the false vacuum}. A stronger criterion is that the physical volume of the false vacuum needs to decrease. This criterion is stronger than the percolation condition since the physical volume of the true vacuum does not only have to increase, it also has to outgrow the expansion of the universe. This condition reads \cite{Turner:1992tz, Ellis:2018mja}
\begin{align}
 \frac1{\mathcal{V}_\text{false}} \frac{\mathrm d \mathcal{V}_\text{false}}{\mathrm d t} = 3 H(t) - \frac{\mathrm d I (t)}{\mathrm d t} = H(T) \left(3 + T \frac{\mathrm d I (T)}{\mathrm d T} \right) < 0\,. 
 \label{eq:completion-PT}
\end{align}
The phase transition completes if this criterion is fulfilled at the percolation temperature. In \cite{Ellis:2018mja}, a regime of even stronger phase transitions was explored. There, this criterion does not hold at the percolation temperature, but instead holds at lower temperatures. Here, we do not make that distinction and only check if \eqref{eq:completion-PT} is fulfilled at $T_p$.

\subsection{Calculation of gravitational-wave spectra}
\label{sec:GW-spectra}
From the mean bubble separation $R$ (or equivalently the inverse duration $\tilde{\beta}$ in cases without strong supercooling), the wall speed $v_w$ and the strength parameter $\alpha$, the GW spectrum at emission can be calculated. Due to the cosmic expansion, the peak frequency at emission, $f_\text{peak}^{\ast}$, is red-shifted to the peak-frequency today, $f_\text{peak}$, according to
\begin{align}
f_\text{peak} = \frac{a_p}{a_0}f_\text{peak}^{\ast}\,,
\end{align}
where $a_p$ is the scale-factor at the time of emission, i.e., at percolation time, and $a_0$ is the scale factor today. Given an expansion history of the universe, the relation between $f_\text{peak}$ and $f_\text{peak}^{\ast}$ can be calculated. For instance, assuming that the universe transitioned into radiation-dominated right after the phase transition and has expanded adiabatically ever since, one obtains \cite{Kamionkowski:1993fg}
\begin{align}
f_\text{peak} =1.65\cdot 10^{-5}\,\text{Hz} \left( \frac{g_*}{100}\right)^{\frac{1}{6}}\left( \frac{T_\text{reh}}{100\,\text{GeV}}\right)\frac{f_\text{peak}^{\ast}}{H(T_p)}.
\end{align}
Here, $g_{\ast}$ is the effective number of relativistic degrees of freedom at the time of production and $T_\text{reh}$ is the temperature at reheating. Using energy conservation $a_\text{reh}^3\rho_\text{rad}(T_\text{reh})=a_p^3\left(\rho_\text{rad}(T_p)+\Delta V\right)$, the reheating temperature can be calculated \cite{Ellis:2018mja},
\begin{align}
T_\text{reh}= \left(\frac{a_p}{a_\text{reh}}\right)^\frac{3}{4}\left( \frac{30\rho_\text{rad}(T_p)}{g_*\pi^2}\left(1+\frac{\Delta V}{\rho_\text{rad}(T_p)}\right)\right)^\frac{1}{4} \simeq T_p \Big(1+\alpha (T_p) \Big)^{\frac{1}{4}} \,.
\label{reheating temperature}
\end{align}
Here, $a_\text{reh}$ is the scale factor at reheating and is assumed to be $a_\text{reh}\simeq a_p$. It holds that $T_\text{reh}\leq T_c$, with the critical temperature $T_c$. For phase transitions that last about a Hubble time or less at a temperature in the 100 GeV range, the peak frequency  falls into the LISA and/or DECIGO sensitivity region.

Once released, the gravitational waves are described in the linearized approximation, such that the total spectrum of gravitational waves is simply the sum of the spectra from three processes, 
\begin{align}
h^2 \Omega_\text{GW} = h^2 \Omega_\text{coll} + h^2 \Omega_\text{sw} + h^2 \Omega_\text{turb} \,.
\end{align}
Herein, $h=H_0/(100\,\text{km/s/Mpc})$ is the dimensionless Hubble parameter today, and the three different $\Omega$ denote the GW amplitudes produced by the  i) collisions of bubble walls~\cite{Kosowsky:1991ua, Kosowsky:1992rz, Kosowsky:1992vn, Kamionkowski:1993fg, Caprini:2007xq, Huber:2008hg, Caprini:2009fx, Espinosa:2010hh, Weir:2016tov, Jinno:2016vai} ($\Omega_\text{coll}$), ii) sound waves in the plasma after bubble collision~\cite{Hindmarsh:2013xza, Giblin:2013kea, Giblin:2014qia, Hindmarsh:2015qta, Hindmarsh:2017gnf}  ($\Omega_\text{sw}$) and iii) magnetohydrodynamic turbulence in the plasma~\cite{Kosowsky:2001xp, Dolgov:2002ra, Caprini:2006jb, Gogoberidze:2007an, Kahniashvili:2008pe, Kahniashvili:2009mf, Caprini:2009yp, Kisslinger:2015hua} ($\Omega_\text{turb}$). We follow \cite{Caprini:2015zlo, Caprini:2019egz, Wang:2020jrd} and use the following expressions to obtain the respective GW spectra from $\alpha$ and $R$:
\begin{align}
 h^2\Omega_\text{coll} (f)&= h^2{\Omega}_\text{coll} ^\text{peak} \frac{3.8( f/f_\text{peak}^\text{coll} )^{2.8}}{1+2.8(f/f_\text{peak}^\text{coll} )^{3.8}}, 
\notag \\[1ex]
 h^2\Omega_\text{sw}(f)&= h^2\Omega_\text{sw}^\text{peak} \left(\frac{f}{f_\text{peak}^\text{sw}}\right)^{3}
\left[ \frac{4}{7}+\frac{3}{7}\left( \frac{f}{f_\text{peak}^\text{sw}} \right)^2\right]^{-\frac{7}{2}}, 
\notag \\[1ex]
 h^2\Omega_\text{turb}(f) &= \frac{h^2{\Omega}_\text{turb}^\text{peak}}{\left( 1+ \frac{8\pi f a_0}{a_p H(T_p)} \right)}\left(\frac{f}{f_\text{peak}^\text{turb}}\right)^3 \left[1+\left( \frac{f}{f_\text{peak}^\text{turb}} \right) \right]^{-\frac{11}{3}}.
\label{GW amplitude}
\end{align}
The peak frequencies read
\begin{align}
f_\text{peak}^\text{coll}  &\simeq 1.65\cdot 10^{-5}\, \text{Hz}\left(\frac{g_*}{100} \right)^{\frac{1}{6}}\left( \frac{T_\text{reh}}{100\,\text{GeV}}\right)\left(\frac{(8\pi)^{\frac 13}}{H(T_p)R} \right)\left(\frac{0.62v_w}{1.8-0.1v_w+v_w^2}\right),
\notag \\[1ex]
f_\text{peak}^\text{sw}&\simeq 1.9\cdot 10^{-5}\,\text{Hz}\left(\frac{g_*}{100} \right)^{\frac{1}{6}}\left( \frac{T_\text{reh}}{100\, \text{GeV}}\right)  \left(\frac{(8\pi)^{\frac 13}}{H(T_p)R} \right),
\notag \\[1ex]
f_\text{peak}^\text{turb}&\simeq 2.7\cdot 10^{-5}\, \text{Hz}\left( \frac{g_*}{100}\right)^{\frac{1}{6}}\left( \frac{T_\text{reh}}{100\, \text{GeV}}\right) \left(\frac{(8\pi)^{\frac 13}}{H(T_p)R} \right).
\label{peak frequency}
\end{align}
For fast phase transitions without strong supercooling, the mean bubble separation $R$ can be approximated by \eqref{mean bubble separation in fist phase transition} for which we obtain the well-known formulae~\cite{Caprini:2015zlo} written in terms of the inverse duration time $\beta$. In a situation with strong supercooling, $R$ is instead given by \eqref{eq:meanbubsepstrongssc}. The peak amplitudes for each spectrum are given by
\begin{align}
 h^2{\Omega}_\text{coll} ^\text{peak}&\simeq 1.67\cdot 10^{-5}\left(\frac{H(T_p)R}{(8\pi)^{\frac 13}}\right)^{2}\left(\frac{\kappa_\text{coll}  \alpha}{1+\alpha} \right)^2\left( \frac{100}{g_*}\right)^{\frac{1}{3}} \frac{0.11v_w}{0.42+v_w^2}\,,
\notag \\[1ex]
 h^2\Omega_\text{sw}^\text{peak}&\simeq 2.65\cdot 10^{-6}\left(\frac{H(T_p)R}{(8\pi)^{\frac 13}}\right)\left( \frac{\kappa_\text{sw} \alpha}{1+\alpha} \right)^2\left(\frac{100}{g_*}\right)^{\frac{1}{3}},
\notag \\[1ex]
 h^2{\Omega}_\text{turb}^\text{peak}& \simeq 3.35\cdot 10^{-4}\left(\frac{H(T_p)R}{(8\pi)^{\frac 13}}\right)\left(\frac{\kappa_\text{turb}\alpha}{1+\alpha}\right)^{\frac{3}{2}}\left( \frac{100}{g_*}\right)^{\frac{1}{3}},
\label{GW spectrum}
\end{align}
where $\kappa_\text{coll}$, $\kappa_\text{sw}$, and $\kappa_\text{turb}$ are the efficiency factors that denote how much $\alpha$ is converted into the energy of the wall (scalar field) and the bulk motion of bubbles. These factors depend on $v_w$ and $\alpha$. We review the detailed formulae in \autoref{App: definitions of efficiencies}. Within the total $\alpha$, the effective quantity $\kappa_\text{coll} \alpha$ is transferred to the strength of gravitational waves produced by the collision of bubbles, and then the remnant $\alpha_\text{eff}=\alpha(1-\kappa_\text{coll})$ becomes an energy source for the dynamics of bubbles generating gravitational waves from sound waves and turbulence. The efficiency factors $\kappa_\text{sw}$ and $\kappa_\text{turb}$ are given by
\begin{align}
\kappa_\text{sw}&= \left( H(T_p) \tau_\text{sw} \right)^{\frac{1}{2}} \kappa_v\,,
&
\kappa_\text{turb}&=\big[1- \left( H(T_p) \tau_\text{sw} \right)\big]^{\frac{2}{3}} \kappa_v\,.
\label{efficiencies for SW and turb}
\end{align}
Here, $\tau_\text{sw}$ is the length of the sound-wave period~\cite{Ellis:2019oqb,Ellis:2020awk} and the efficiency factor $\kappa_ v$ is the fraction of the energy transferred into the bulk motion of bubbles. The longer the sound-wave period lasts, the less energy remains available for GW production through turbulence, accordingly $\kappa_\text{turb}$ decreases with increasing $\tau_\text{sw}$. The length of the sound-wave period is defined by
\begin{align}
\tau_\text{sw}= \text{min}\!\left[\frac{1}{H(T_p)}, \frac{R_*}{\bar U_f}\right].
\label{the length of the sound wave period}
\end{align}
Thus, $H(T_p) \tau_\text{sw}$ takes into account the reduction of GW spectrum from sound waves when the sound-wave period, in which $h^2\Omega_\text{sw}$ is actively produced, is shortened~\cite{Ellis:2018mja}. The length of the sound-wave period \eqref{the length of the sound wave period} depends on the root-mean-square fluid velocity of the plasma~\cite{Hindmarsh:2015qta,Ellis:2019oqb},
\begin{align}
\bar U_f^2 = \frac{3}{v_w(1+\alpha)}\int^{v_w}_{c_s}\!\mathrm d\xi \, \xi^2 \frac{v(\xi)^2}{1-v(\xi)^2}\simeq  \frac{3}{4}\frac{\alpha_\text{eff}}{1+\alpha_\text{eff}}\kappa_v,
\end{align}
where $c_s=1/\sqrt{3}$ is the speed of sound, $v(\xi)$ is the solution to \eqref{differential equation for the plasma velocity profile}, and we assumed $v_w\simeq 1$ in the second equality. The efficiency factor $\kappa_ v$ depends on the velocity of the bubble wall, see \eqref{general efficiency factor kappaV} in \autoref{App: definitions of efficiencies}. For $v_w=1$, we have
\begin{align}
\kappa_v=\frac{\alpha_\text{eff}}{\alpha}\frac{\alpha_\text{eff}}{0.73+0.083\sqrt{\alpha_\text{eff}}+\alpha_\text{eff}}.
\end{align}
Several comments are in order: Firstly, the formulae for the GW amplitudes depend on underlying assumptions, e.g., for the bubble dynamics. In particular, the formulae \eqref{GW amplitude} have been derived in the envelope approximation in which colliding bubble walls immediately lose their energy, see, e.g., \cite{Weir:2016tov, Jinno:2016vai}. The validity of this approximation, however, strongly depends on the features of both the phase transition and the bubble dynamics, especially on the magnitude of $\alpha$. One can distinguish two cases for the bubble dynamics before collisions. One is the runaway bubble case, in which the phase transition produces a lot of energy, i.e., a large $\alpha$, most of which is transferred to the bubble wall. This case, in which incidentally the envelope approximation breaks down, requires a huge $\alpha$ of the order of $10^{12}$~\cite{Bodeker:2017cim}, which is far beyond the range of $\alpha$ achievable in our setup. The other is the non-runaway bubble case. The bubble wall reaches a constant terminal velocity due to the pressure and the friction with the plasma. In this case, one typically relies on the envelope approximation, as we will do here, although it has recently been questioned whether the envelope approximation is valid for $\alpha\gg1$ \cite{Jinno:2019jhi}. An improvement of the envelope approximation for GWs from bubble collision was for example discussed in \cite{Lewicki:2020jiv}. Furthermore, the recent work \cite{Guo:2020grp} has discussed the length of the sound-wave period in an expanding universe: the resulting GW spectrum from sound waves \eqref{GW amplitude} is suppressed by the factor $\Upsilon=1-1/\sqrt{1+2H(T_p)\frac{R_*}{\bar U_f}}$. We do not include this factor here, since its effects are approximately included in the suppression due to the length of the sound-wave period given in \eqref{the length of the sound wave period}.

Secondly, in the non-runaway case, the efficiency factor $\kappa_\text{coll}$ is negligible in comparison with $\kappa_\text{sw}$ and $\kappa_\text{turb}$, \cite{Alanne:2019bsm,Ellis:2019oqb}, which we have checked numerically. Therefore, $\kappa_\text{coll}\approx 0$ is a good approximation for this case. In this paper, we set the bubble-wall velocity to be the speed of light, i.e. $v_w=1$, and use the approximation $\kappa_\text{coll}\approx 0$. We caution that for smaller bubble-wall velocities, a significant suppression (of the order of $10^{-3}$) of the GW amplitude from sound waves has been suggested to occur in comparison with that predicted in earlier works~\cite{Cutting:2019zws}.

\subsection{Summary of the computation of the gravitational-wave spectra}
\label{sec:short-summary}
In summary, we are using the following equations to determine the GW spectra from the effective scalar potential
\begin{itemize}
 \item We use the strength parameter, where the bag constant is related to the trace of the energy-momentum tensor, see \eqref{basic quantity: alpha: free energy}, evaluated at the percolation temperature.
 \item For the mean bubble separation, we distinguish between the cases with and without strong supercooling. For strong supercooling, a minimization temperature exists and then the inverse duration is given by \eqref{eq:beta-strong-supercooling} and the mean bubble separation follows from \eqref{eq:meanbubsepstrongssc}. If no minimization temperature exits, we are in a regime without strong supercooling and the defining equations are \eqref{eq:LinearBeta2} and \eqref{mean bubble separation in fist phase transition}.
 \item For the efficiency parameters, we use $\kappa_\text{coll}\approx 0$ as well as \eqref{efficiencies for SW and turb}, i.e., we include a suppression factor that takes into account the length of the sound-wave period. Furthermore, we set the bubble-wall velocity to be the speed of light $v_w=1$.
 \item The GW spectra are finally determined by \eqref{GW amplitude} -- \eqref{GW spectrum}.
\end{itemize}

\section{Results: Gravitational wave spectra from new-physics contributions in the Higgs potential}
\label{sec:results}
In \autoref{sec:GWcalculation}, we have reviewed the calculation of GW spectra from a first-order electroweak phase transition with a given effective potential. In the present work, we utilize rather generic effective potentials obtained from integrating-out quantum and thermal effects below a given high-energy (TeV) cutoff scale $M_\text{NP}$, see \autoref{sec:HOHiggs}. The initial effective potentials at the ultraviolet scale $M_\text{NP}$ stem from integrating out BSM physics above this scale, and the different classes of potentials  parameterize a wide range of BSM models. 

\begin{figure}[!t]
 \centering
 \includegraphics[width=.65\linewidth]{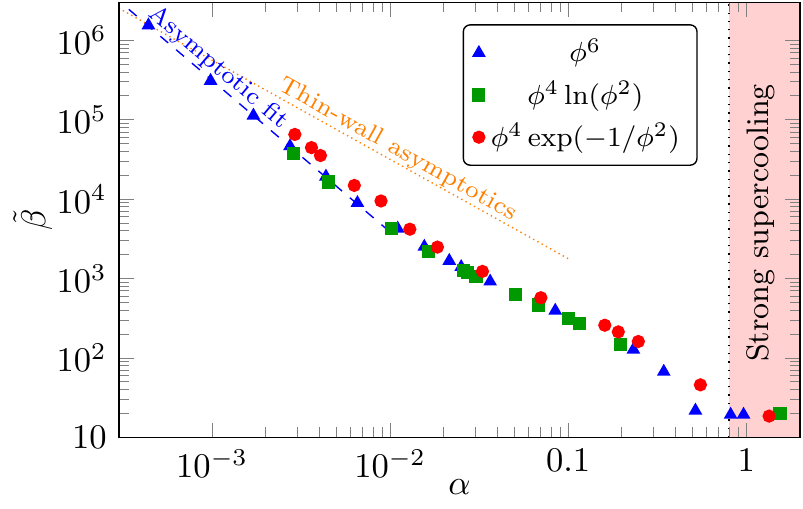}
 \caption{We display the $\alpha$-$\tilde\beta$--values that can be achieved from the three different modifications of the Higgs potential that we consider here. We find an almost universal relation between $\tilde{\beta}$ and $\alpha$ for all modifications. The red area marks the strong supercooling regime as defined in \autoref{sec:strong-supercooling}. At small $\alpha$, we find an asymptotic power-law behavior with $\tilde{\beta}=0.58\cdot \alpha^{-1.91}$. It does not match the asymptotic power-law behavior expected from the thin-wall approximation, see \autoref{App:ThinWall}.}
 \label{fig:BetaAlphaPlane}
\end{figure}

\subsection{Universal GW-parameters from BSM physics}
\label{sec:universality}
The GW spectra are computed from the parameters $\alpha$ and  $R$ (or $\tilde \beta$), that relate to the energy released during the phase transition and the mean bubble separation (or the inverse  time duration), respectively. For the evaluation of potential new physics, it is important to know which combinations of pairs $(\alpha, \tilde \beta)$ can be achieved. The values of ($\alpha,\tilde\beta)$ depend on the specifics of the effective potential around the phase-transition temperature. Varying these potentials freely leads to a scatter plot in the $\alpha$-$\tilde\beta$--plane, see, e.g., \cite{Caprini:2019egz}. 

\begin{figure}[t]
 \includegraphics[width=.99\linewidth]{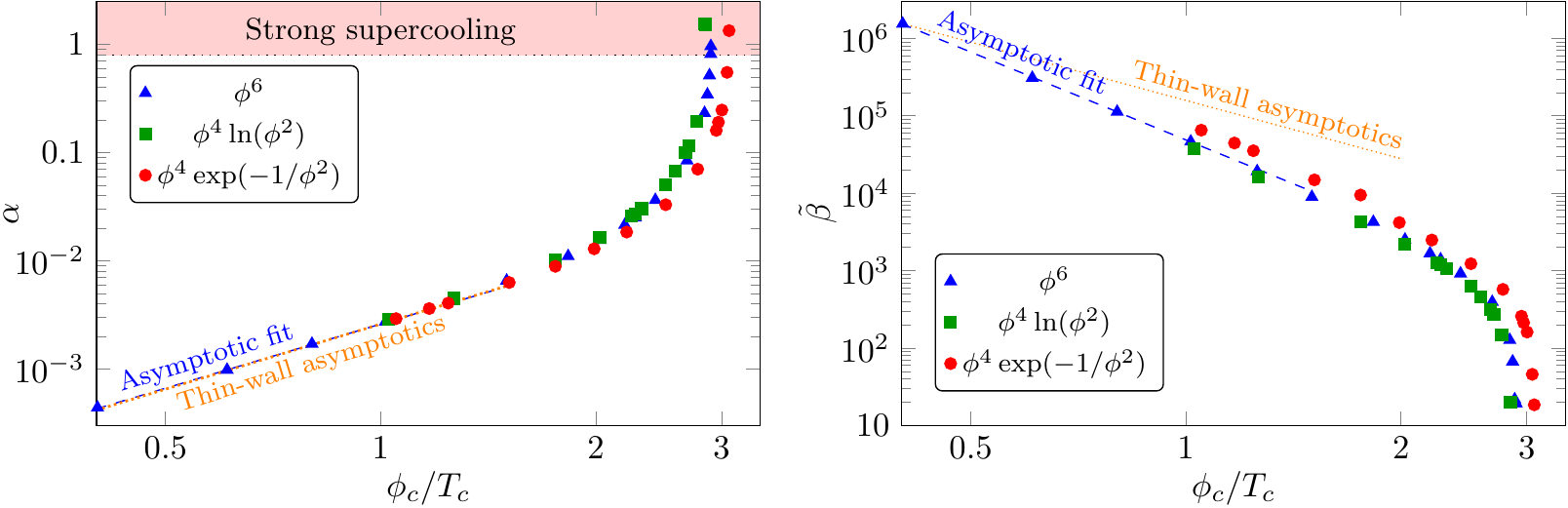}
 \caption{We display $\alpha$ (left panel) and $\tilde\beta$ (right panel) as a function of $\phi_c/T_c$ for all potentials. We find an almost universal relation between $\alpha$,$\tilde{\beta}$ and $\phi_c/T_c$ for all modifications. The red area marks the strong supercooling regime as defined in \autoref{sec:strong-supercooling}. At small $\phi_c/T_c$, we find asymptotic power-law behaviors with $\alpha=0.0026\cdot \left(\phi_c/T_c\right)^{1.97}$ and $\tilde{\beta}=48800\cdot (\phi_c/T_c)^{-3.77}$. The thin-wall approximation matches the power-law behavior of $\alpha$ perfectly, while it does not match that for $\beta$, see \autoref{App:ThinWall}.}
 \label{fig:AlphaBeta-pctc}
\end{figure}

In our study, this scatter plot reduces to a rather small band of parameters in the $\alpha$-$R$ and $\alpha$-$\tilde\beta$--plain, see \autoref{fig:HRAlphaPlane} and \autoref{fig:BetaAlphaPlane}, respectively. This is a remarkable result in view of the wide range of BSM physics our study is expected to cover, and the corresponding diverse shapes of the effective potential. This \textit{universal} curve  arises from corresponding curves for $\alpha$ and $\tilde\beta$ as functions of the vacuum expectation value $\phi_c/T_c$, see \autoref{fig:AlphaBeta-pctc}: with increasing strength of the phase transition encoded in $\phi_c/T_c$, $\alpha$ grows, while $\tilde{\beta}$ decreases. The $\phi_c/T_c$-dependence of $\alpha, \tilde{\beta}$ already shows a roughly universal behavior for the different types of BSM-modifications. Moreover, for small $\phi_c/T_c\lesssim 1$, both $\log \alpha$ and $\log{\tilde{\beta}}$ show a linear dependence on $\log \phi_c/T_c$, indicating a universal power law. In turn, for $\phi_c/T_c\gtrsim  1$, no simple power law is present but universality still holds true.  

The above observations entail that the generic BSM physics encoded in the different potential classes introduced in \autoref{sec:HOHiggs} lead to a universal relation between the energy released during the phase transition and its inverse duration time. For the universality seen in \autoref{fig:HRAlphaPlane}, \autoref{fig:BetaAlphaPlane}, and \autoref{fig:AlphaBeta-pctc}, one would expect the specifics of the initial effective potentials at the UV scale  $M_{\rm NP}$ be washed-out by quantum and thermal fluctuations below the UV scale. Indeed, in the perturbative regime, the RG flow typically washes out the ``memory" of the initial conditions at the cutoff scale $M_{\rm NP}$ over just a few orders of magnitude in scales. Specifically, despite being non-polynomial around the origin in field space, all potentials can be expanded about a given \textit{finite} field value or background $\bar\phi$. The expansion coefficients of $(\phi-\bar\phi)^n$ with $n\geq 6$  exhibit a power-law dependence on the RG-scale $k$, driving them towards zero with $k^{n-4}$. Based on this observation, one might conclude that the underlying reason for our discovery of a universal curve in the $\alpha$-$R$-plane (or $\alpha$-$\tilde{\beta}$-plane) is the underlying universality of the effective potential that emerges from different classes of microphysics. 

\begin{figure}[!t]
 \centering
 \includegraphics[width=0.65\linewidth]{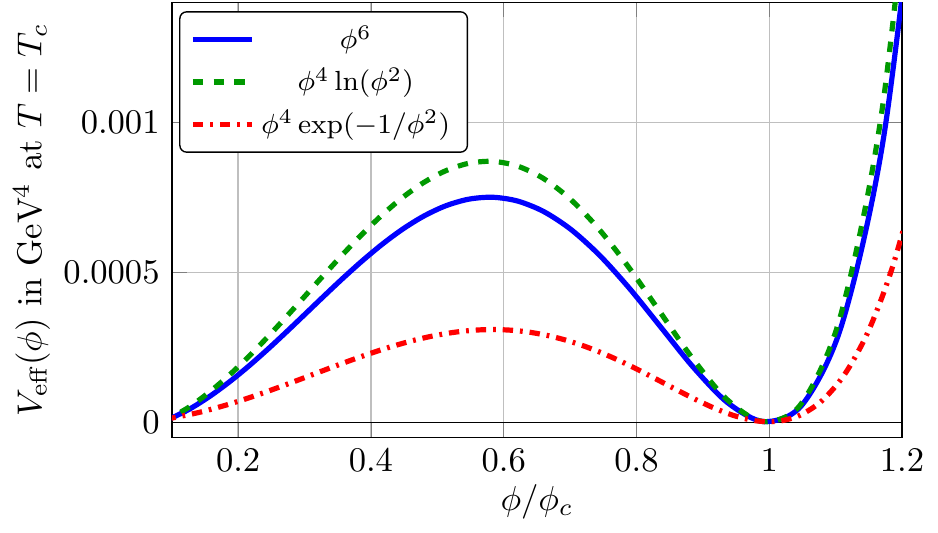}
 \caption{Displayed are the effective potentials for all three modifications of the Higgs potential at $T=T_c$ with $\phi_c/T_c\approx 1$. Despite their differences, the potentials lead to very similar values of $\alpha$ and $R$ and thus also to similar GW spectra. This makes the observed universality in \autoref{fig:HRAlphaPlane}, \autoref{fig:BetaAlphaPlane}, and \autoref{fig:AlphaBeta-pctc}, highly non-trivial. Importantly, the potentials differ in their explicit values of $\phi_c$ and $T_c$: $\phi_{c,\phi^6}=116.2$, $\phi_{c,\ln}=116.4$, $\phi_{c,\exp}=110.5$.}
 \label{fig:effpot}
\end{figure}

We confront this expectation with our numerical data on the finite-temperature effective potential at the critical temperature, which we display in \autoref{fig:effpot}. All potentials displayed in \autoref{fig:effpot} lead to similar values of $\alpha$ and $R$ (or $\tilde \beta$) and they also fulfill $\phi_c/T_c\approx 1$. Nevertheless, the effective potentials exhibit clear differences between the three classes of modifications. These differences between the classes exist at \emph{all} temperatures. At vanishing temperature, they give rise to differences in the three- and four-Higgs coupling, which we have discussed in \autoref{sec:HOHiggs} and displayed in \autoref{fig:Higgs-coupling}. Furthermore,  each potential displayed in \autoref{fig:effpot} has a different value of $\phi_c$, $T_c$, $T_p$, etc., although they have the value of $\phi_c/T_c\approx 1$ in common. Accordingly, the source of the universality of the $\alpha$-$R$ and $\alpha$-$\tilde{\beta}$-relation is not due to a universal form of the effective potential. Instead, it is a form of universality that arises in the integrated information from the effective potential, which enters the GW spectra. Therefore, we expect that there is a universal potential that could be modeled such as to provide the $\alpha(\phi_c/T_c)$, $R(\phi_c/T_c)$,  and $\tilde{\beta}(\phi_c/T_c)$ curves. Such a universal potential would encode only the coarse-grained information reflected in these two parameters. On the other hand, it would not encode the LHC observables, as these distinguish between the three classes of microphysics.

Finally, to investigate the origin of the power-law dependence of $\alpha(\phi_c/T_c)$, $R(\phi_c/T_c)$, and $\tilde\beta(\phi_c/T_c)$, we discuss the regime of $\phi_c/T_c\lesssim 1$. For very small $\phi_c/T_c$, we expect to be able to rely on the thin-wall approximation with $\epsilon(T_p)=V_\textrm{eff}(0,T_p)- V_\textrm{eff}(\langle \phi\rangle_{T_p},T_p)\to 0$. In this limit, close to the second-order phase transition, analytic computations are accessible, see \autoref{App:ThinWall}: the percolation temperature $T_p$ is close to the critical temperature and we can expand all quantities in the reduced temperature $\delta_c(T) =(T_c-T)/T_c$. Concentrating on the leading coefficients, this leads to scaling relations for $\alpha$ and $\tilde\beta$ as a function of $\phi_c/T_c$, which we displayed in \autoref{fig:AlphaBeta-pctc}. The power-law for $\alpha$ matches the observed asymptotic fit very well, while we note a clear difference in the power-law for $\tilde \beta$. We conjecture that the reason is that we have not yet reached small enough values of $\phi_c/T_c$, such that the thin-wall approximation is sufficiently fulfilled. Furthermore, the leading-order coefficient of the expansion in $\tilde \beta$ may be strongly numerically suppressed, which is why we instead observe the next-to-leading order behavior. Importantly, the $\alpha$ relation does not depend on the thin-wall approximation, but on the $\phi^4$ approximation, which explains the excellent matching in the $\alpha$ relation, see \autoref{App:ThinWall} for more details. We leave a more detailed study that could explain the quantitative values of both power laws to future studies.

\begin{figure}[!t]
 \includegraphics[width=.99\linewidth]{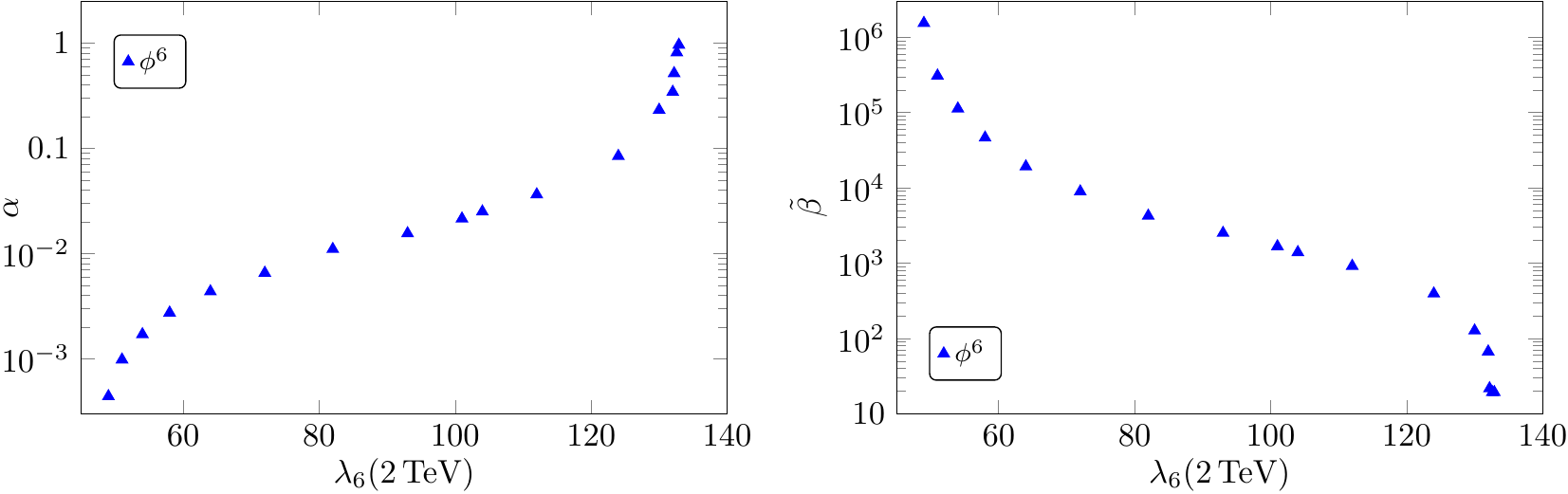}
 \caption{We display $\alpha$ (left panel) and $\tilde \beta$ (right panel) as a function of $\lambda_6$ at the new-physics scale $M_\text{NP}=2$\,TeV. These relations can be used to directly match a given new-physics model to the GW signature.}
 \label{fig:alpha-beta-lam6}
\end{figure}

In summary, we make the tentative discovery that, unlike light new physics, heavy new physics in the TeV range leads to a universal $\alpha(\tilde{\beta})$ curve. This has ramifications for both GW detections as well as collider searches, that we will spell out in detail below.

\subsection{Model building}
To map perturbative new-physics models directly to GW-parameters, we provide \autoref{fig:alpha-beta-lam6}, which links the $\phi^6$ coupling at the new-physics scale to GW parameters. For this class of new physics, this coupling is sufficient to capture the salient effect of the new physics. In terms of GW physics, all modifications at the new-physics scale effectively reduce to one universal potential with the universal relations between $\alpha$ and $R$ ($\tilde \beta$) as shown in \autoref{fig:HRAlphaPlane} and \autoref{fig:BetaAlphaPlane}. The universality of our results indicates that a family of universal effective potentials should exist, which captures the GW parameters correctly. This family of potentials depends on the parameters $\phi_c$, $T_c$, and $T_p$. For these potentials, we relate the parameters of the effective field theory description at the cutoff scale $M_{\rm NP}$ to the parameters $\alpha$ and $\tilde\beta$, cf.~\autoref{fig:alpha-beta-lam6}. This is useful in terms of model-building, as it provides a more direct map from new-physics models to the GW signature: In a given perturbative new-physics model, one can calculate the size of the $\phi^6$ coupling at the cutoff scale $M_{\rm NP}$. Within a perturbative new-physics setting, the $\phi^6$ coupling is expected to capture the leading-order contribution. In \cite{Chala:2018ari} it has been shown that already the next-to-leading order $\phi^8$ contribution has a strongly subleading effect on the phase transition. The size of $\lambda_6$ in turn translates directly into corresponding GW parameters. We have compared our results from the $\phi^6$ modification of the Higgs potential with \cite{Leitao:2015fmj, Ellis:2018mja, Chala:2018ari, Wang:2020jrd}. In this comparison, we focused on physical information such as the phase transition temperature $T_p$, the strength parameter $\alpha$, and the inverse duration time of the phase transition $\tilde \beta$, while we gave less importance to scheme dependent quantities such as the new-physics scale. Within the physical phase transition parameters, our results were in qualitative agreement with \cite{Leitao:2015fmj, Ellis:2018mja, Chala:2018ari, Wang:2020jrd}.

In summary, our results here provide a direct map from perturbative new physics, which can be captured in terms of a $\phi^6$ coupling, to the GW parameters. This provides a direct way to estimate, for a given perturbative new-physics model, whether or not it is likely to be detectable by GW detectors.

\begin{figure}[t]
	\centering
	\includegraphics[width=7.5cm]{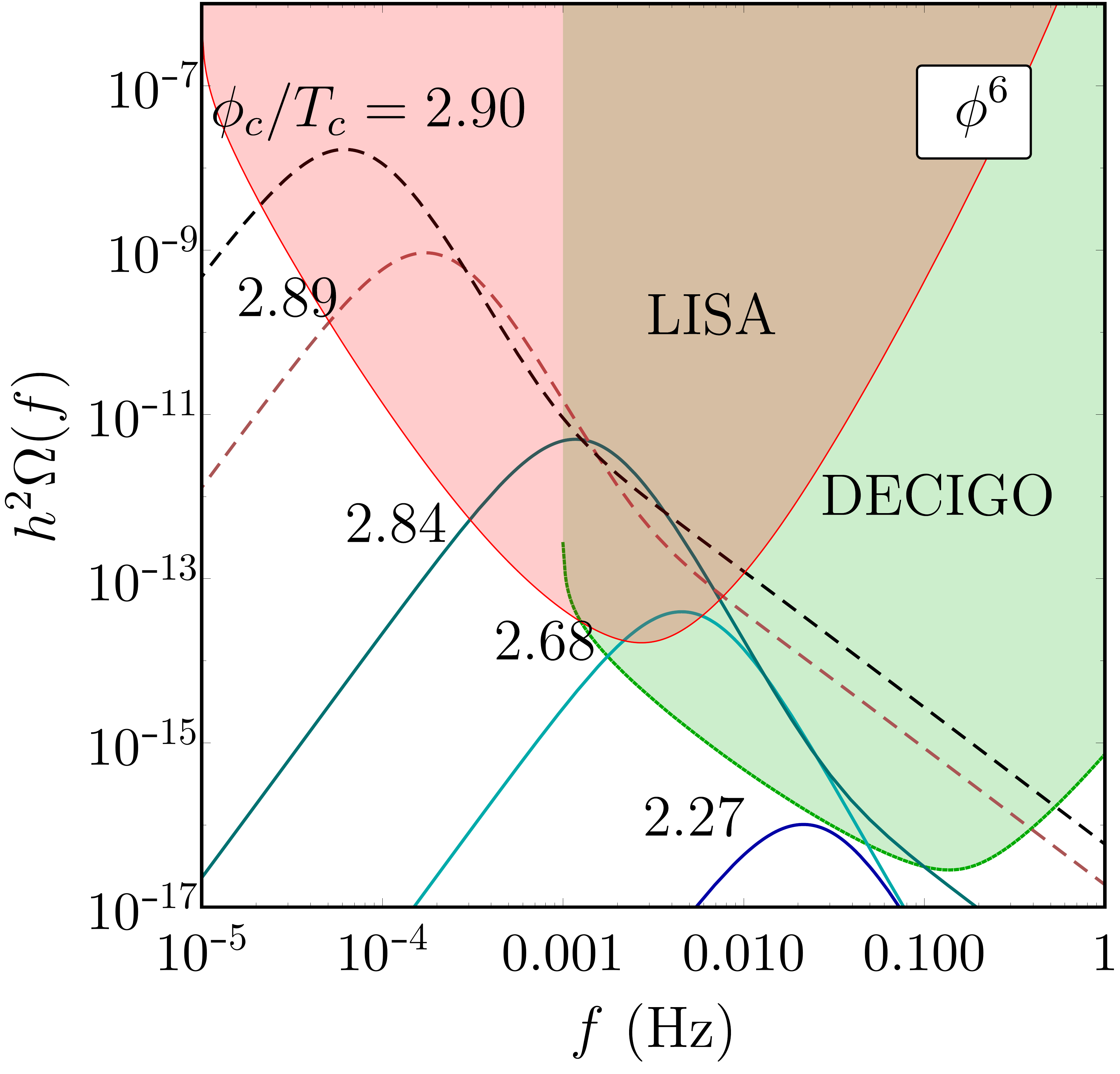}
	\hfill
	\includegraphics[width=7.5cm]{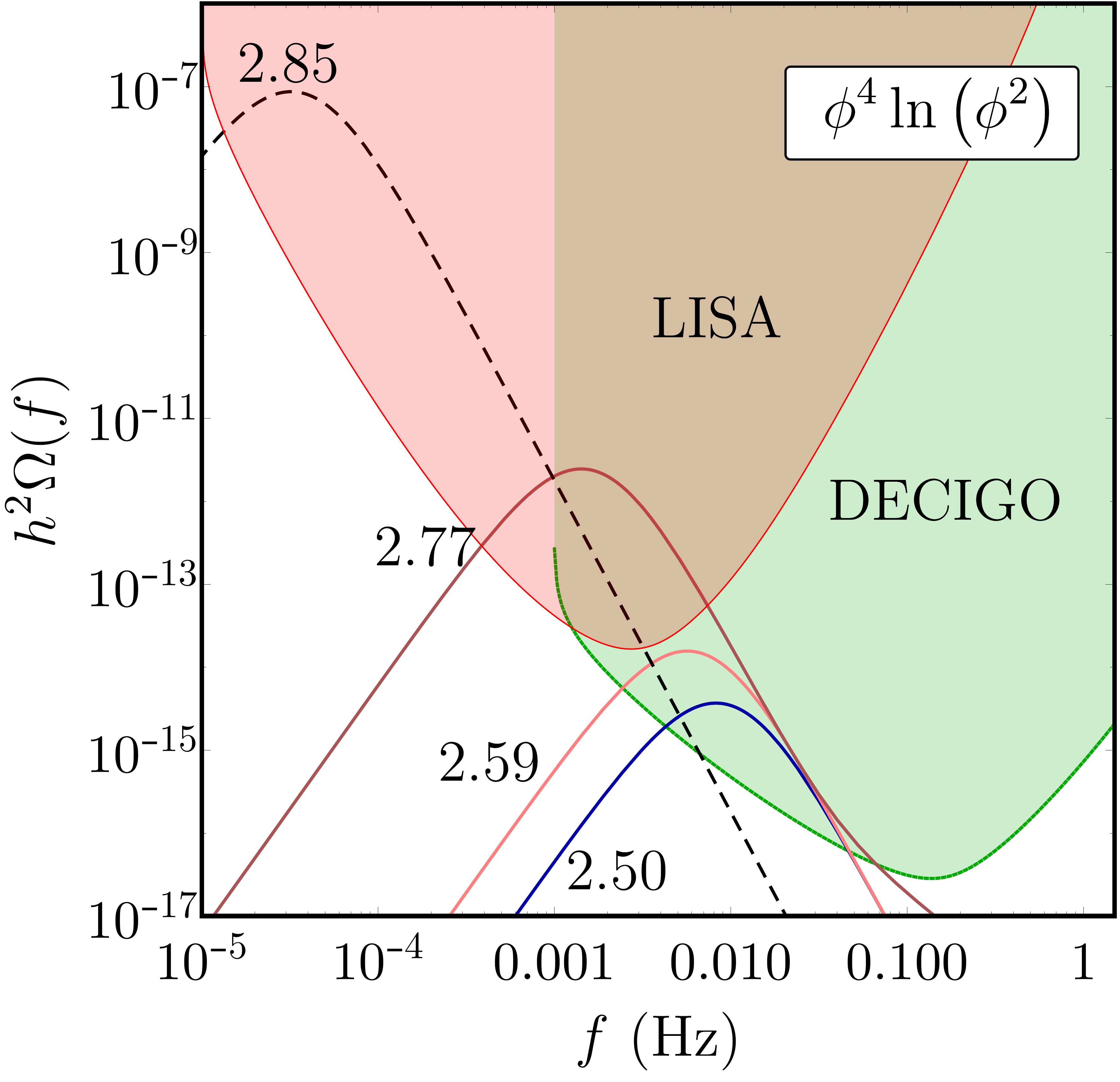}\\[1ex]
	\includegraphics[width=7.5cm]{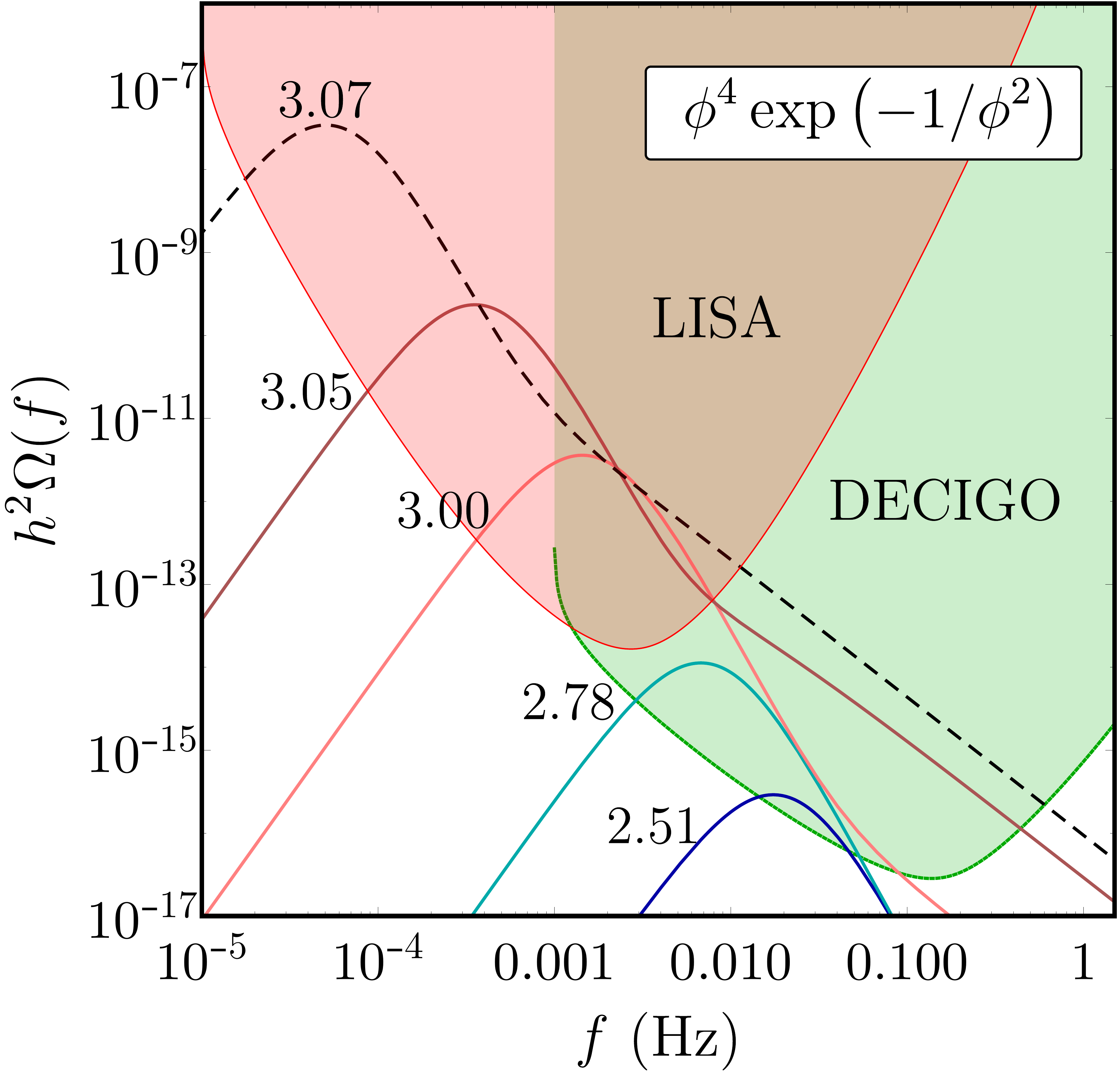}
	\caption{We display the GW spectra for all modifications of the Higgs potential for different values of $\phi_c/T_c$. The top-left panel displays the polynomial modifications, the top-right panel the logarithmic modifications, and the bottom panel the exponential modifications.  The dashed GW spectra are in the strong supercooling regime as defined in \autoref{sec:strong-supercooling}. }
	\label{fig:SpectrumAll} 
\end{figure}

\subsection{Gravitational-wave spectra}
\label{sec:gw-spectra}
We compare the GW spectra for the three classes of potentials to the LISA and DECIGO power-law integrated sensitivity curves, taken from \cite{LISAdocument, Seto:2001qf, Kawamura:2006up, Kawamura:2011zz}, a nice overview over the power-law integrated sensitivity curves of GW detectors is given in \cite{Schmitz:2020syl}. Our first result, in agreement with the literature, is the detectability of the GW signal at both of these planned detectors for electroweak phase transitions which are strong enough, cf.~\autoref{fig:SpectrumAll}. Phase transitions with $\phi_c/T_c \approx 2 - 2.7$ lie inside the LISA sensitivity band. Therefore, phase transitions with a smaller $\phi_c/T_c$ would require an instrument at higher frequencies and increased sensitivity, such as DECIGO. The alternative technology underlying AION and AEDGE \cite{Badurina:2019hst}, using atom interferometry, might  be able to reach sensitivities relevant for an electroweak phase transition of lower $\phi_c/T_c$. In particular, our results indicate that LISA is mostly sensitive to phase transitions with strong supercooling. In contrast, DECIGO is also sensitive to those phase transitions with moderate supercooling.

As a general trend, increasing strength of the phase transition, i.e., larger $\phi_c/T_c$ results in a shift of the peak-frequency towards lower frequencies, as well as a growth of the amplitude. The same trend has already been observed, e.g., for the class of $\phi^6$ potentials, see, e.g., \cite{Leitao:2015fmj}. Due to the universality of $\alpha({\tilde{\beta}})$, this feature is shared by all three classes of potentials, cf.~the three panels in \autoref{fig:SpectrumAll}. Due to the shift in frequency with increasing $\phi_c/T_c$, the peak frequency of the GW spectra lies in the outer areas of the LISA , as well as DECIGO sensitivity bands, for those cases where the maximum amplitude lies above the sensitivity curves, respectively. 

\begin{figure}[t]
	\centering
	\includegraphics[width=.47\linewidth]{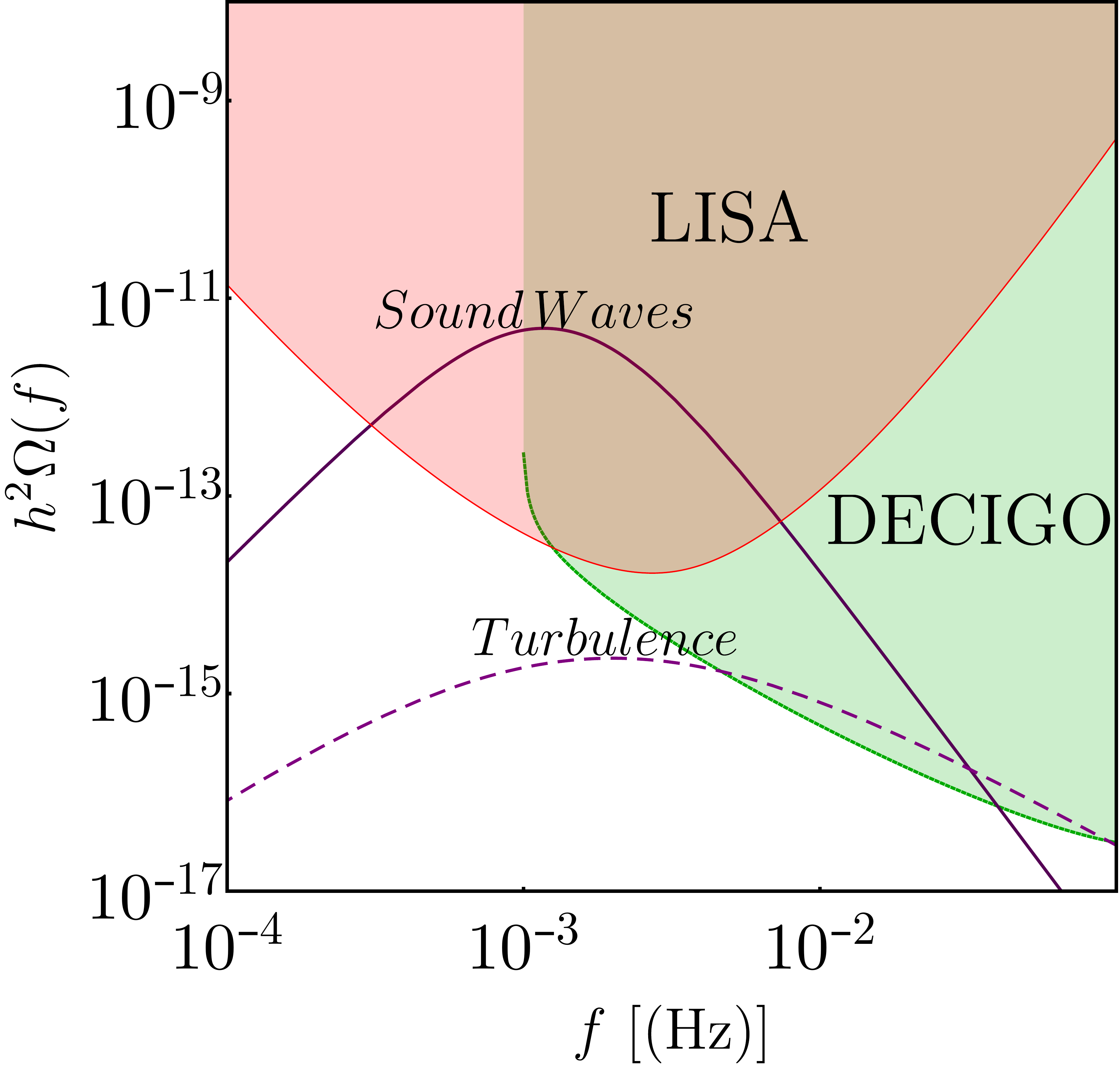} \hfill
	\includegraphics[width=.5\linewidth]{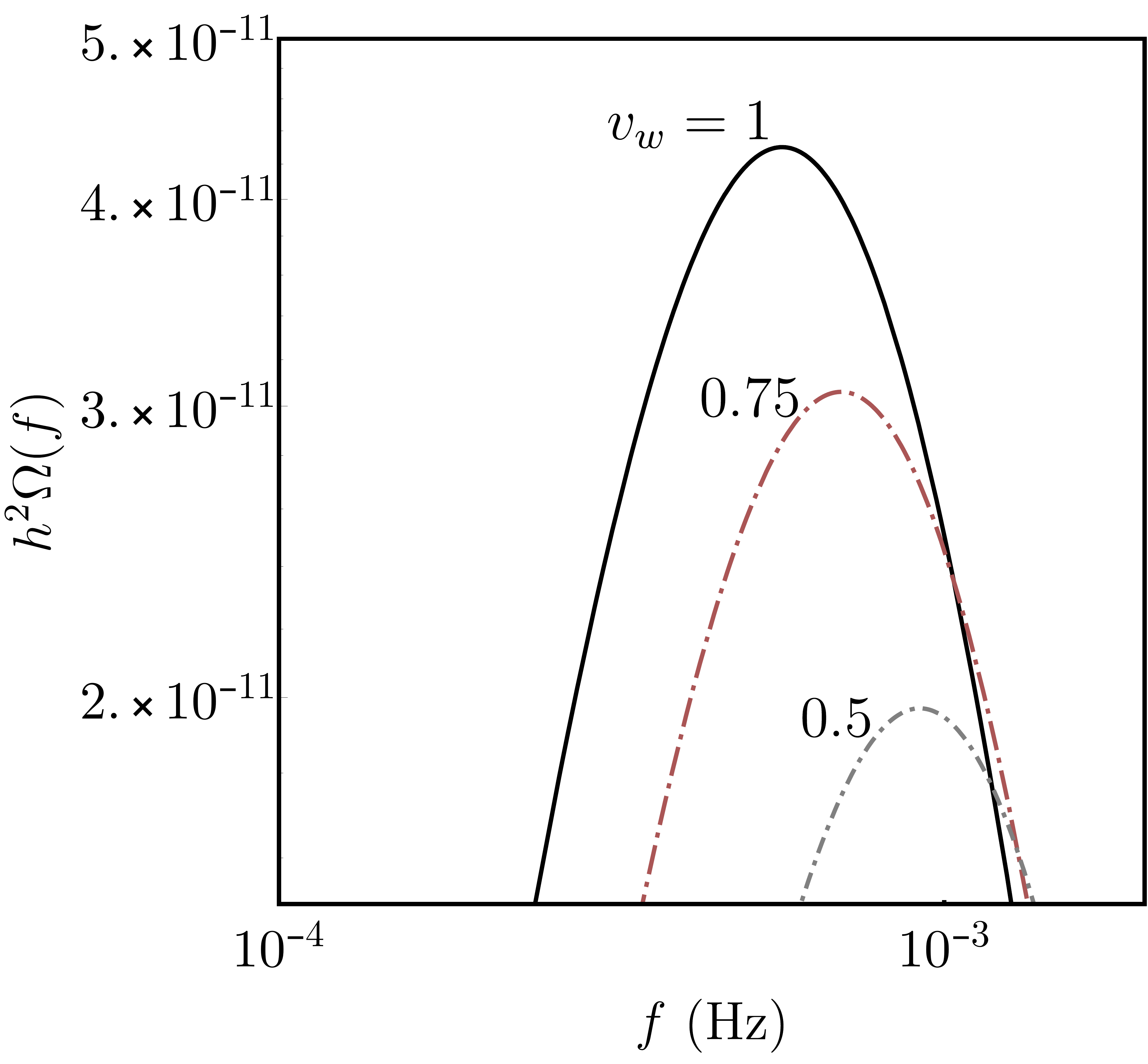}
	\caption{\textbf{Left panel:} We show the contributions of sound waves and turbulence to the GW spectrum. The sound-wave contribution is the dominant contribution for all investigated cases. \textbf{Right panel:} We display the effect of the wall speed on the GW spectrum. With increasing wall speed the spectra are shifted towards larger amplitudes and peak frequencies. The effect is small compared to changes in $\phi_c/T_c$.  For both panels, the displayed example spectrum stems from the polynomial modification of the Higgs potential with $\phi_c/T_c=2.86$.}
	\label{fig:amplitudes-wallspeed}
\end{figure}

The GW spectra depend on several parameters, such as the efficiency factors that determine how much energy is converted into gravitational waves, or the wall speed of the expanding bubbles. Our assumptions and choices for these parameters are summarized in \autoref{sec:short-summary}. As is well-known in the literature, the sound-wave contribution is typically the dominant component of the spectrum, see \cite{Hindmarsh:2015qta} for results from numerical simulations. As a self-consistency check of our treatment, our results pose no exception, as is exemplified in the left panel of \autoref{fig:amplitudes-wallspeed}. We have confirmed that the same is true for all other cases we analyze. Note that our use of \eqref{efficiencies for SW and turb} for the efficiency factors is the most conservative choice, as it leads to a lower amplitude for the dominant sound-wave contribution than other choices that can also be found in the literature.

Further, we test the robustness of our results under variations of the wall speed, which we set to $v_w=1$ for the main part of this work. Decreasing it generically results in a lower amplitude and slightly larger peak frequencies, and the difference between $v_w=0.5$ and $v_w=1$ can easily be a factor of 2 in peak amplitude, cf.~the right panel of \autoref{fig:amplitudes-wallspeed}. Our choice of the maximum wall speed is accordingly a less conservative one.

For GW detectors, the relevant quantity that informs about the detectability of a stochastic GW signal is the signal-to-noise-ratio (SNR) \cite{Allen:1997ad, Maggiore:1999vm}, which can be obtained from the GW spectrum, the sensitivity curve of the detector, $h^2 \Omega_{\rm det}$, and the observation time $T$, obtained from the duration of the mission times the duty cycle, as in \cite{Thrane:2013oya}  
\begin{align}
\text{SNR} = \sqrt{\frac{T}{s} \int_{f_{\rm min}}^{f_{\rm max}}\mathrm df \left(\frac{h^2 \Omega_{\rm GW}}{h^2 \Omega_{\rm det}} \right)^2}. 
\end{align}
Typically, an observation time of roughly four years is assumed, with a duty cycle of 75\% \cite{LISAdocument} i.e., $T \approx 3\pi \cdot 10^7$\,s. We emphasize that $h^2 \Omega_{\rm det}$ is the sensitivity curves of the detector, not to be confused with the power-law integrated sensitivity curves displayed in \autoref{fig:SpectrumAll} and \autoref{fig:amplitudes-wallspeed}. To determine a threshold SNR is not straightforward, as the detectability of a signal is influenced by various aspects, such as, e.g., whether matched filtering techniques are applicable which in general improves the detectability of a signal significantly \cite{LISAdocument}. In our case, as the spectrum is known, these techniques are applicable. Additionally, galactic binaries constitute an expected stochastic background-signal that needs to be accounted for \cite{Adams:2010vc, Adams:2013qma}. We obtain SNRs very significantly above 1 for cases where the peak amplitude at peak frequency lies above the LISA sensitivity curve. Let us note that we used $f_\text{min}=10^{-4}$\,Hz as a lower cutoff for the sensitivity of LISA, in agreement with \cite{LISAdocument}. For the phase transitions with strong supercooling, the SNR would be significantly increased by an extension to, e.g., $f_\text{min}=10^{-3}$\,Hz, see \autoref{fig:SpectrumAll}.

In summary, we conclude that heavy new physics could leave detectable imprints at future GW detectors. For LISA, cases without strong supercooling are more challenging to access, calling for a GW observatory with increased sensitivity. In contrast to light new physics, where a stronger GW signal has been predicted in many cases, the characteristic shift in peak-frequency towards lower frequencies that heavy-new-physics models exhibit, makes it more challenging to detect their imprints.

\begin{figure}[t]
 \includegraphics[width=0.99\textwidth]{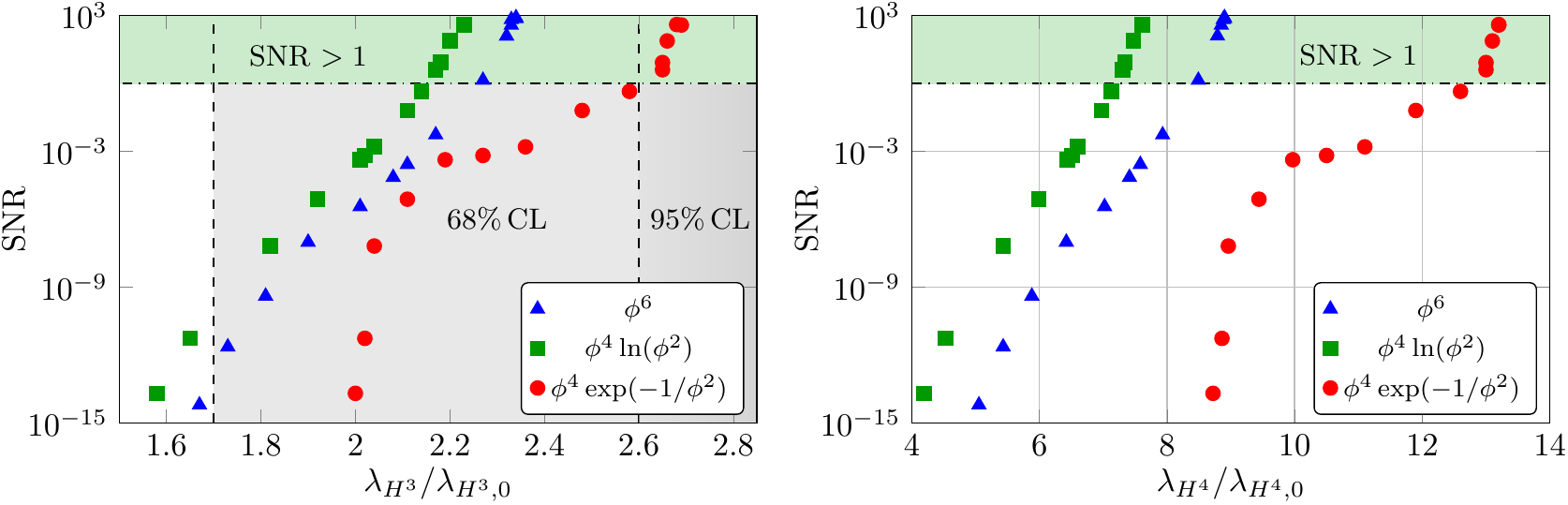}
 \caption{We display the normalized three-Higgs (left panel) and four-Higgs self coupling (right panel) as a function of the LISA SNR for the three different modifications of the Higgs potential. The green area indicates a SNR above 1 and the grey areas display the confidence level with which the high-luminosity run of the LHC can exclude the modification of the Higgs self-coupling, see \eqref{eq:lhc}. A combined effort of  GW detectors and particle collider is needed to distinguish between the different models of NP.}
 \label{fig:Higgs-SNR}
\end{figure}

\subsection{Collider vs gravitational-wave signatures}
\label{sec:GWandLHC}
In this section, we explore how GW and collider signatures can be used concertedly to learn about new physics. In particular, we focus on aspects of universality and how to distinguish different types of new-physics contributions.

The new physics that triggers a strong first-order phase transition and thereby a GW signal, at the same time affects observable properties of the Higgs potential, namely the effective three-Higgs and four-Higgs coupling, see \autoref{fig:Higgs-coupling}. Both are enhanced over their values in the SM without new physics, as has been investigated in \cite{Reichert:2017puo}. The high-luminosity run of the LHC will be able to test deviations in the Higgs self-couplings. The standard channel to measure the three-Higgs coupling at the LHC is Higgs pair production in gluon fusion \cite{Eboli:1987dy, Dicus:1987ic, Glover:1987nx, Plehn:1996wb, Djouadi:1999rca, Baur:2002rb, Baur:2002qd, Baur:2003gpa, Baur:2003gp, Baglio:2012np, Li:2013rra}, see, e.g., the diagrams in Fig.\,1 of \cite{Reichert:2017puo}. An exact cancellation occurs in the SM in the low-energy kinetic regime, leading to a distinct signal, should the SM relation be violated. The optimal reach of the high-luminosity LHC run with $3\,\text{ab}^{-1}$ is given by~\cite{Kling:2016lay},
\begin{align}
\frac{\lambda_{H^3}}{\lambda_{H^3,0}} &= \phantom{-} 0.4~...~1.7\, \qquad \text{at 68\% CL\,,} \notag\\[1ex]
\frac{\lambda_{H^3}}{\lambda_{H^3,0}} &= -0.2~...~2.6\, \qquad \text{at 95\% CL\,.}
\label{eq:lhc}
\end{align}
A value for $\lambda_{H^3}/\lambda_{H^3,0}$ outside the given range would detectably violate the cancellation of di-Higgs production in gluon fusion low-energy kinetic regime as discussed above.

The cosmological constraint that the electroweak phase transition must complete, as discussed in \autoref{sec:max-strength}, has an intriguing implication for the Higgs self-couplings observed at the LHC: it yields a maximal possible value of the three and four-Higgs coupling, assuming that no other mechanism, i.e., light degrees of freedom, modify the Higgs self-coupling. We find that these maximal values are given by $\lambda_{H^3,\text{max}}/\lambda_{H^3,0}= 2.69$ and $\lambda_{H^4,\text{max}}/\lambda_{H^4,0}= 13.2$.

In \autoref{fig:Higgs-SNR}, we compare the LHC observables against the SNR for the GW signal for all three classes of potentials. As expected, strong enough phase transitions are expected to lead to detectable GW signals at LISA as well as a detectable LHC signal. Therefore, the two observational signatures can be used as cross-checks of each other: For instance, an enhancement of the triple-Higgs and quartic-Higgs coupling might come from other types of new physics, which leave the cross-over intact, and therefore do not provide the conditions for electroweak baryogenesis. To strengthen the case for new physics that changes the cross-over to a strong first-order phase transition requires the observation of a GW signal. Conversely, GW signals could also arise, e.g., from first-order phase transitions in a dark sector \cite{Schwaller:2015tja}. The LHC constraints on the three- and four-Higgs coupling could distinguish this from a phase transition in the electroweak sector.

Moreover, even amongst those scenarios with electroweak baryogenesis, a combination of the GW signal with the LHC signature might allow learning more about the details of the new physics. \autoref{fig:Higgs-SNR} highlights that at a given GW SNR, the different classes of potentials lead to different signal strengths at the LHC. In some cases, these might potentially even be distinguishable. 

\section{Conclusions and outlook}
\label{sec:conclusions}
In this paper, we have explored the GW signal sourced by a first-order electroweak phase transition arising from BSM physics. As in \cite{Reichert:2017puo}, the new physics is parameterized by an addition to the effective potential at the new-physics scale $M_\text{NP}$. We focus on ``heavy" new physics, $M_\text{NP}=2$\,TeV, such that there are no additional light BSM degrees of freedom. We have performed a non-perturbative analysis, by integrating out quantum and thermal fluctuations below the scale of new physics to obtain the finite-temperature effective potential. From the latter, we extract key parameters that determine the GW spectrum.

The key novel result of our study is an unexpected emergence of universality: We observe that the energy released by the phase transition, $\alpha$, and the mean bubble separation, $R$, as well as the inverse duration of the phase transition, $\beta$, show a qualitatively universal dependence on $\phi_c/T_c$ for the distinct types of new-physics contributions we have explored, see \autoref{fig:AlphaBeta-pctc}. This results in a clustering of $R(\alpha)$ and $\beta(\alpha)$ around a common curve, see \autoref{fig:HRAlphaPlane} and \autoref{fig:BetaAlphaPlane}. Such a behavior is quite surprising because we investigated rather distinct possibilities for the new physics by exploring a wide range of new-physics contributions at the new-physics scale. Specifically, we investigate an effective-field-theory inspired $\phi^6$ contribution, a Coleman-Weinberg-inspired logarithmic contribution, and a non-perturbative exponential contribution. We believe these to cover most generic cases of new-physics contributions with a new-physics scale $\gtrsim 1\,\rm TeV$ with no new degrees of freedom below that scale. In fact, not only the UV potentials but also the corresponding effective potentials at and below the phase-transition temperature differ. Consequently, LHC observables, such as the three-Higgs self-coupling, do not exhibit universality. Hence the emergence of universality for the $R(\alpha)$ curve is nontrivial. The universality holds for new-physics scales significantly above the electroweak scale, i.e., heavy new physics. Our result should be contrasted with the situation where the new-physics contribution is associated with a lower mass scale, and more of the $\alpha-R$ plane is accessible when model parameters are varied. Indeed, many examples in the literature illustrate that models with light new degrees of freedom generically deviate from our newly-discovered universal curve, see, e.g., the examples in \cite{Caprini:2019egz}.

It is an intriguing future goal to understand the roots of this qualitative universality in more detail, including a classification of potential, more exotic new-physics classes that are not captured by our universal result. Our results suggest that a universal effective potential exists that is determined by a single free parameter in addition to depending on the phase-transition temperature $T_p$ and critical field value $\phi_c$, which encodes the impact of heavy new degrees of freedom on the GW parameters. We stress that such a universal effective potential may be used to extract the GW parameters, but cannot account for the particle-physics observables, where non-universal features of the distinct classes of new physics matter. 

We expect this result to have ramifications for GW searches as it severely constrains the available parameter space for GW signatures that arise from heavy new physics -- a property that was not previously noticed in the literature to the best of our knowledge.

Our results have direct relevance for the phenomenology of GW observations. Firstly, our results severely constrain the available parameter space for GW signatures -- a property that was not previously noticed in the literature to the best of our knowledge. Secondly, observational results that do not lie on the universal curve are a strong indication for the presence of light new degrees of freedom. Thirdly, under the hypothesis that our findings are representative of many (if not all) new-physics contributions with a high new-physics scale, the maximum achievable SNR of a GW signal would follow from our study. We find an SNR at LISA that suggests that some new-physics cases should be detectable. At DECIGO, an even larger range of detectable phase transitions opens up.

Besides GW observatories, collider experiments, e.g., the LHC, could be sensitive to such new-physics contributions, through a deviation of the effective three-Higgs and four-Higgs couplings from their SM- values. As the expected signal could be challenging to detect both at the LHC as well as at most GW observatories; a convincing detection could be more within reach with the use of both types of instruments. Our discovery of universality strengthens the case for a combination of both signatures: As a consequence of universality for the heavy new physics, GW signals could provide a way to generically distinguish heavy from light new physics: if a signal is detected that cannot be constructed by values of $\alpha$ and $\beta$ on the universal curve, this is a strong indication for the presence of light new physics. In contrast, distinguishing information on the form of the heavy new physics is not encoded in the GW signal, i.e., the GW signals for different models are degenerate. Interestingly, as shown in \cite{Reichert:2017puo}, the strength of the LHC-signal differs for the different classes of new-physics contributions. The LHC might be able to lift the degeneracy of GW signatures of distinct new physics, as the same value of $\alpha(R)$ (and therefore the GW signal) is associated with different values of the effective Higgs couplings in different models. A concerted effort, involving both GW observatories as well as the LHC, therefore seems indicated. 

GWs are already used together with electromagnetic and neutrino signals as part of an ongoing effort to constrain fundamental physics with multi-messenger astronomy. Here, we highlight that another promising, multi-instrument campaign, using GW observatories together with the LHC, might be yet another way in which fundamental physics beyond the SM might be constrained or even discovered.

\subsection*{Acknowledgements}
We acknowledge helpful discussions with Ryusuke Jinno.
A.E.~and J.L.~acknoweldge support by an Emmy-Noether grant of the DFG under grant number Ei/1037-1. A.E.~is supported by a Villum Young Investigator grant of VILLUM FONDEN under grant no.~29405.
J.M.P.~is supported by the DFG Collaborative Research Centre SFB 1225 (ISOQUANT) and the DFG under Germany's Excellence Strategy EXC - 2181/1 - 390900948 (the Heidelberg Excellence Cluster STRUCTURES).
M.R.~is supported by the Science and Technology Research Council (STFC) under the Consolidated Grant ST/T00102X/1.
The work of M.Y.~is supported by an Alexander von Humboldt Fellowship. 

\appendix

\section{Details of the Functional Renormalization Group approach}
\label{App:FRG}
With the functional renormalization group, one initiates the flow at a UV scale, in our case the new-physics scale $M_\text{NP}$, and successively integrates out quantum and thermal fluctuations without relying on perturbation theory. This yields the temperature-dependent effective Higgs potential, from which we can read off properties such as the vacuum expectation values (vev) $v$, the Higgs mass $m_H$, the Higgs self-couplings $\lambda_{H^3}$ and $\lambda_{H^4}$, as well as the order and strength of the phase transition $\phi_c/T_c$. The former quantities are read off from the zero-temperature potential, while the latter follows from the temperature-dependent one. 

In our functional renormalization group approach, we are working in the framework of \cite{Eichhorn:2015kea}, i.e., we do not work in a fully fledged SM. Instead we only implement the flow equations of the $SU(3)$ gauge coupling $g_3$, a fiducial coupling $g_F$ that simulates the $SU(2)$ and the $U(1)$ sector, the top-Yukawa coupling $y_t$, and the full Higgs potential $V(\phi)$. Accordingly, instead of a complex $SU(2)$ doublet, we focus on a real scalar for the Higgs field. It has been confirmed in \cite{Eichhorn:2015kea} that this approximation suffices to evaluate the Higgs potential with good accuracy. The flow of the Higgs potential is evaluated numerically on a grid in the field $\phi$. This allows us to implement non-perturbative modifications of the Higgs potential, such as $\Delta V(\phi) = \frac14 \lambda_{\ln} \phi^4\ln \frac{\phi^2}{2 M_\text{NP}^2}$ and $\Delta V(\phi) = \frac18 \lambda_{\exp} \phi^4 \exp\!\left(-\frac{2 M_\text{NP}^2}{\phi^2} \right)$, see \eqref{eq:mod-Higgs-pot}. All flow equations at finite temperature are detailed in \cite{Reichert:2017puo}.

The new parameters parameterizing the modification of the Higgs potential (i.e., $\lambda_6$, $\lambda_{\ln}$, and $\lambda_{\exp}$, see \eqref{eq:mod-Higgs-pot}) allow us to control the strength of the phase transition $\phi_c/T_c$. For each $\phi_c/T_c>0$, there is a corresponding $\lambda_i>0$ at $M_\text{NP}$.  When changing the new-physics contribution $\Delta V$, we adjust $\mu$, $\lambda_4$, and $y_t$ at the initial scale $M_\text{NP}$ such that the following low-energy values remain unchanged:
\begin{align}
	v &= 246 \,\text{GeV}\,,
	&
	m_H &= 125 \,\text{GeV}\,, 
	&
	m_t &= 173 \,\text{GeV} \,,
	\label{eq:ir_data}
\end{align}
where the top mass is obtained from the top-Yukawa coupling together with the vev. In practice, adjusting $\mu$, $\lambda_4$, and $y_t$ at the initial scale $M_\text{NP}$ such that \eqref{eq:ir_data} holds is a tuning-problem for the initial conditions and can be implemented numerically. As long as $M_\text{NP}$ is not too large ($M_\text{NP}\lesssim 10$\,TeV, see \cite{Reichert:2017puo}), we can always find values of $\mu$, $\lambda_4$, $y_t$, and $\lambda_i$ at $M_\text{NP}$ that lead to \eqref{eq:ir_data} together with our desired strength of the phase transition. 

In this paper, we set $M_\text{NP} =  2$\,TeV throughout but we expect our results to hold at least in the range 1\,TeV\,$\lesssim M_\text{NP} \lesssim 10$\,TeV. We emphasize that only at the initial scale $M_\text{NP}$ the Higgs potential is given by the analytic form in \eqref{eq:mod-Higgs-pot}. Below the initial scale, we take all contributions into account and the potential does not possess a simple form anymore. 

\begin{figure}[t]
	\includegraphics[width=0.99\textwidth]{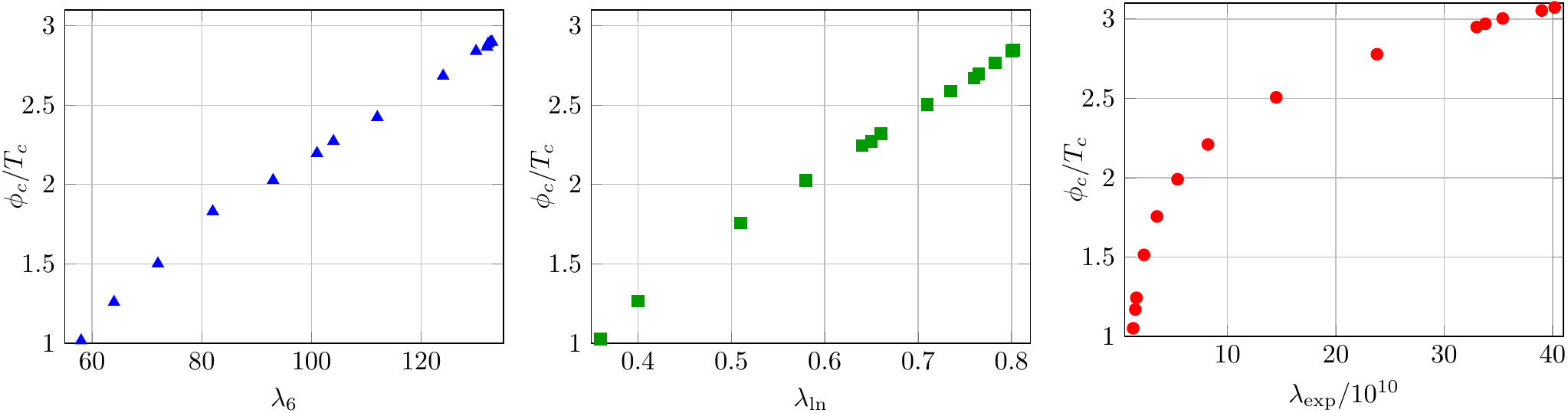}
	\caption{Order parameter of the phase transition $\phi_c/T_c$ as a function of the parameters of the higher-order modifications of the Higgs potential at a new-physics scale $M_\text{NP} = 2$\,TeV. Note that the parameter $\lambda_{\exp}$ is rather large due to the strong suppression of the exponential factor.}
	\label{fig:pctc}
\end{figure}

The values of $\lambda_i$ for which the modifications of the Higgs potential lead to a strong first-order phase transition are given in \autoref{fig:pctc} as well as in the result tables in \autoref{app:res-tables}. Note that $\lambda_{\exp}$ is rather large due to the strong suppression of the exponential factor.

\section{Efficiencies of gravitational-wave production processes}
\label{App: definitions of efficiencies}
The GW spectra depend on efficiency factors that describe the fractions of vacuum energy transferred into the dynamics of bubbles, see \eqref{GW spectrum}. The efficiency factor $\kappa_\text{coll}$ denotes the fraction stored in the bubble wall, which is released upon collisions of the bubble walls, while the energy converted into bulk motion of the fluid is characterized by $\kappa_\text{sw}$ and $\kappa_\text{turb}$, respectively.

For $\kappa_\text{coll}$, we need to understand the dynamics of the bubble wall, which we briefly review now. Imposing the thin-wall approximation, the Lagrangian of the bubble radius $r=r(t)$ is given by~\cite{Darme:2017wvu}
\begin{align}
\mathcal{L} =- M_\text{wall}(r)\gamma +\frac{4\pi}{3}r^3 p\,.
\label{Eq: Lagrangian of bubble radius}
\end{align}
Here $M_\text{wall}(r)=4\pi r^2 \sigma$ can be regarded as the ``mass" of the bubble wall, $\gamma=\sqrt{1-\dot r^2}$ is the Lorentz factor and $p$ is the pressure on the bubble wall. The mass of the bubble wall is proportional to the bubble-wall tension $\sigma$, which is identified with the one-dimensional Euclidean action $S_1$, see \eqref{S1 action in the thin-wall approximation} in \autoref{App:ThinWall}. The dot on $r$ denotes the time derivative, i.e., $\dot r={\mathrm d}r/{\mathrm d}t$. The equation of motion for $r$ is given by
\begin{align}
\frac{{\mathrm d} \gamma}{{\mathrm d} r} + \frac{2\gamma}{r}=\frac{p}{\sigma}\,.
\end{align}
With the initial condition $\gamma(r_0)=1$, the solution is 
\begin{align}
\gamma = \frac{p}{3\sigma}r +\frac{r_0^2}{r^2} -\frac{p}{3\sigma}\frac{r_0^3}{r^2} \,.
\label{Eq: Lorentz factor}
\end{align}
From \eqref{Eq: Lagrangian of bubble radius}, the total energy of the bubble wall is given by
\begin{align}
E_\text{tot}= (4\pi   r^2 \sigma )\gamma -\frac{4\pi}{3}r^3 p\,.
\label{Eq: energy of bubble wall}
\end{align}
The critical radius $r_c$ at which the total energy is minimized, i.e., ${\mathrm d}E_\text{tot}/{\mathrm d}r=0$, is given by $r_c=2\sigma/p$ under the assumption of $\gamma=1$ and constant $p$ and $\sigma$ under variations of $r$. In order that the bubble can expand, its initial radius has to be larger than $r_c$. Assuming that the initial radius is slightly larger than the critical one, i.e., $r_0 \gtrsim r_c$, the Lorentz factor can be approximated by $\gamma \approx \frac{2}{3}\frac{r}{r_0} +\frac{1}{3}\frac{r_0^2}{r^2}$. As the bubble radius increases,  the second term is suppressed, whereas the first term dominates. Therefore, the approximated relation between $\gamma$ and $r$ reads
\begin{align}
\gamma \approx \frac{2}{3}\frac{r}{r_0}\,.
\label{Eq: approximated Lorentz factor}
\end{align}
One can see from the Lagrangian \eqref{Eq: Lagrangian of bubble radius} that the pressure on the bubble wall is a key quantity for the bubble dynamics. The relevant contribution that causes the expansion of the bubble comes from the vacuum energy difference between the two vacua as given in \eqref{energy density of vacuum}. The friction force between the bubble wall and plasma can contribute to a reduction of the pressure. In general, it is complicated to evaluate the friction force. The highly relativistic case allows us to approximately obtain the pressure on the bubble wall:
\begin{align}
p\simeq \Delta V_\text{eff} -\Delta P_\text{LO} -\gamma \Delta P_\text{NLO}\,,
\label{Eq: total pressure on the bubble wall}
\end{align}
with 
\begin{align}
\Delta P_\text{LO}&=\frac{\Delta m^2 T^2}{24}\,,
&
\Delta P_\text{NLO} &= \frac{g^2\Delta m_V T^3}{24}\,.
\end{align}
Here $\Delta P_\text{LO}$ and $\Delta P_\text{NLO}$ are the leading-order and next to leading-order pressures due to the friction induced by $1\to1$ and $1\to2$ scattering processes across the bubble wall. The term $\Delta P_\text{LO}$ describes the $1\to1$ processes and  is proportional to the squared-mass difference between the symmetric and broken phases, $\Delta m^2 = \sum_i c_i N_i \Delta m_i^2$, weighted by  the number of degrees of freedom for particle species $i$, $N_i$, and factors $c_i=1$ for a boson and $c_i=1/2$ for a fermion~\cite{Bodeker:2009qy}. The term $\Delta P_\text{NLO}$ describes the $1\to2$ processes, e.g., $e^{-}\to e^{-}Z$~\cite{Bodeker:2017cim}.  $\Delta P_\text{LO}$ is frame independent, i.e., it does not depends on the Lorentz factor, while the contribution from $1\to2$ scattering processes is proportional to $\gamma$.\footnote{The recent work \cite{Hoeche:2020rsg} has improved the computation given in \cite{Bodeker:2017cim} and has discussed that the next to leading-order pressure should be given by $\gamma^2\Delta P_\text{NLO}\propto \gamma^2T^4$ rather than $\gamma\Delta P_\text{NLO}\propto \gamma\Delta m_V T^3$. Using this result, the efficiency factor $\kappa_\text{coll}$ has been also improved in \cite{Ellis:2020nnr}.} The $1\to 2$ processes yield the mass difference, $g^2\Delta m_V = \sum_i g_i^2 N_i \Delta m_i$, weighted by the gauge coupling constant $g_i$.  Although one can consider, in general, the $1\to 2$ processes via the scalar self-coupling and the Yukawa interactions, the dominant contributions come from the longitudinal vector boson emission $\phi\to V_L \phi$ where $\phi$ and $V_L$ denote a SM particle and the longitudinal mode of a massive vector boson ($W^{\pm}$ or $Z$), respectively.

The treatment \eqref{Eq: total pressure on the bubble wall} is only valid when $\Delta V_\text{eff}>\Delta P_\text{LO}$. As $\gamma$ increases, we expect that the pressure \eqref{Eq: total pressure on the bubble wall} vanishes, $p=0$, when the Lorentz factor reaches
\begin{align}
\gamma_\text{eq}= \frac{\Delta V_\text{eff} -\Delta P_\text{LO}}{\Delta P_\text{NLO}}\,,
\end{align}
at which a bubble has the radius $r_\text{eq}=\left(3\gamma_\text{eq}/2 \right)r_0$ from \eqref{Eq: approximated Lorentz factor}. The Lorentz factor $\gamma$ becomes constant once it reaches $\gamma_\text{eq}$. Thus, the acceleration of the bubble wall stops and it expands with a constant velocity. In other words, the frame-independent net pressure (vacuum energy) $\Delta V_\text{eff} -\Delta P_\text{LO}$, which accelerates the bubble wall, balances with the next to leading-order friction $\gamma\Delta P_\text{NLO}$ at $\gamma=\gamma_\text{eq}$.

The efficiency factor $\kappa_\text{coll}$ describes how much of the vacuum energy is used to accelerate the bubble wall. The total vacuum energy is simply given by $\Delta V_\text{eff}$ times the volume of the bubble at percolation, i.e., $E_V=\Delta V_\text{eff} \frac{4\pi}{3}r_*^3$, where we denoted the bubble radius at percolation by $r_*$. For the bubble-wall energy, we distinguish whether the bubble radius reaches $r_\text{eq}$ before percolation or not. As long as $r<r_\text{eq}$, the bubble wall accelerates and thus all vacuum energy, except for the leading-order friction, is transferred into the acceleration of the bubble wall. Thus, if $r_*<r_\text{eq}$ (or equivalently if $\gamma_*<\gamma_\text{eq}$) then the bubble-wall energy is given by$E_\text{wall}=(\Delta V_\text{eff} -\Delta P_\text{LO}) \frac{4\pi}{3}r_*^3$. Then, the efficiency factor $\kappa_\text{coll}$ is given by
\begin{align}
\kappa_\text{coll} &=\frac{E_\text{wall}}{E_V} = \frac{\Delta V_\text{eff} -\Delta P_\text{LO}}{\Delta V_\text{eff}}
&
(r_*&<r_\text{eq})\,. 
\label{Eq: kappa collision for r*<req}
\end{align}
For $r>r_\text{eq}$, the bubble wall does not accelerate anymore and the bubble radius grows with a constant rate. Thus, in the case of  $r_*>r_\text{eq}$ the efficiency factor reads~\cite{Ellis:2019oqb}
\begin{align}
\kappa_\text{coll} &=\frac{E_\text{wall}}{E_V} = \frac{(\Delta V_\text{eff} -\Delta P_\text{LO}) \frac{4\pi}{3}r_\text{eq}^3}{\Delta V_\text{eff}\frac{4\pi}{3}r_*^3} + \frac{\Delta A}{\Delta V_\text{eff}\frac{4\pi}{3}r_*^3} \,,
&
(r_* &> r_\text{eq})\,.
\label{Eq: kappa collision for r*>req}
\end{align}
where the first term accounts for the transferred energy before $r_\text{eq}$ and the second term for the transferred energy after $r_\text{eq}$. The numerator in the second term is the kinetic energy in \eqref{Eq: energy of bubble wall} between $r_\text{eq}\leq r\leq r_*$, given by
\begin{align}
\Delta A=4\pi (r_*^2 - r_\text{eq}^2)\sigma_\text{eq} \gamma_\text{eq} = \frac{4\pi}{3} \Delta V_\text{eff}(r_*^2 - r_\text{eq}^2) r_\text{eq}\,,
\label{Eq: bubble area difference}
\end{align}
with $\sigma_\text{eq}=r_\text{eq}\Delta V_\text{eff}/(3\gamma_\text{eq})$ the bubble wall tension at $r_\text{eq}$ which is derived from the first term in \eqref{Eq: Lorentz factor} only, as the last two terms are suppressed in this case.
The quantity \eqref{Eq: bubble area difference} entails the increase in the bubble-wall area when the bubble expands with a constant velocity.

We rewrite these results in more convenient dimensionless quantities given by
\begin{align}
\alpha_\infty&=\frac{\Delta P_\text{LO}}{\rho_\text{rad}}\,,
&
\alpha_\text{eq}&=\frac{\Delta P_\text{NLO}}{\rho_\text{rad}}\,,
\end{align}
where $\alpha_\infty$ denotes the weakest phase transition for which the vacuum pressure exceeds the leading-order friction. $\alpha_\text{eq}$ is defined such that we can rewrite the terminal Lorentz factor with $\gamma_\text{eq} = \frac{\alpha-\alpha_\infty}{\alpha_\text{eq}}$. Furthermore, we define $\gamma_*=\frac{2}{3}\frac{r_*}{r_0}$, which is the Lorentz factor that the bubble wall would reach if the next-to-leading-order friction was neglected. The condition $r* \lessgtr r_\text{eq}$ is then equivalent to $\gamma_* \lessgtr \gamma_\text{eq}$. The efficiency factor for the collision of bubbles, \eqref{Eq: kappa collision for r*<req} and \eqref{Eq: kappa collision for r*>req}, can be written as
\begin{align}
\kappa_\text{coll}= \begin{cases}
\frac{\gamma_\text{eq}}{\gamma_*} \left[ 1 -\frac{\alpha_\infty}{\alpha} \left(\frac{\gamma_\text{eq}}{\gamma_*} \right)^2 \right], & \gamma_*>\gamma_\text{eq},
 \\[1ex]
1-\frac{\alpha_\infty}{\alpha},  &  \gamma_* \leq \gamma_\text{eq}.
\end{cases}
\label{efficiency factor for collision}
\end{align}
The remnant $\alpha_\text{eff}=\alpha(1-\kappa_\text{coll})$ is transferred into the bulk motion of bubbles which produces GW from sound wave and turbulence. Note that if the vacuum energy dominates over the friction, $\kappa_\text{coll}$ tends to be unity and then $\alpha_\text{eff}\simeq 0$, so that GW from the bulk motion of bubbles are suppressed.

The efficiency factors for sound waves and turbulence, $\kappa_{\rm sw}$ and $\kappa_{\rm turb}$, are determined in terms of the efficiency factor $\kappa_v$, cf.~\eqref{efficiencies for SW and turb}. The efficiency factor $\kappa_v$ is defined by~\cite{Espinosa:2010hh}
\begin{align}
\kappa_v(\alpha, v_w)= \frac{3}{\rho_\text{vac}v_w^3} \int^{v_w}_{c_s} w(\xi)\, \xi^2 \frac{v(\xi)^2}{1-v(\xi)^2}\mathrm d\xi,
\label{efficiency factor kappav}
\end{align}
where $c_s=1/\sqrt{3}$ is the speed of sound.  The integral accounts for all velocities within the bubble, which are limited by the speed of sound on the one side and the wall velocity on the other side; where both $c_s>v_w$ or $c_s<v_w$ are possibly, see below.
$w(\xi)$ is the plasma enthalpy profile,
\begin{align}
w(\xi)= w_0 \exp\!\left[ (1-c_s^{-2}) \int^{v(\xi)}_{v_0} \mathrm d v' \frac{\mu(\xi,v')}{1-v'{}^2} \right].
\end{align}
Here, we defined the Lorentz transformed fluid velocity 
\begin{align}
\mu(\xi,v)=\frac{\xi-v}{1-\xi v},
\end{align}
and the plasma velocity profile $v(\xi)$, which is obtained from the differential equation
\begin{align}
\frac{2v}{\xi}=\frac{1-\xi v}{1-v^2}\left[ c_s^{-2}\mu^2(\xi,v)-1 \right] \frac{\partial v}{\partial \xi}.
\label{differential equation for the plasma velocity profile}
\end{align}
Depending on the initial conditions, the solutions to \eqref{differential equation for the plasma velocity profile} are classified into three different types~\cite{Espinosa:2010hh}: (i) deflagrations, (ii) detonations and (iii) hybrids.
(i): The bubble-wall velocity is subsonic, i.e., $v_w<c_s$. Therefore, collisions between the bubble wall and the fluid outside are mostly avoided and little energy goes into the generation of sound waves and turbulence. In this case, the GW signal from sound waves tends to be suppressed~\cite{Cutting:2019zws}.
(ii): The bubble wall velocity is supersonic ($v_w>c_s$). The active fluid is inside the wall, while the wall hits the fluid at rest.
Hence, the strong gravitational waves from sound waves and turbulence could be produced. 
(iii): The bubble wall moves at a supersonic speed, but smaller than the Chapman-Jouguet detonation velocity $v_J$ which is defined by
\begin{align}
v_J=\frac{\sqrt{2\alpha_\text{eff}/3+\alpha_\text{eff}^2}+\sqrt{1/3}}{1+\alpha_\text{eff}}.
\end{align}
This case is a combination of the two cases (i) and (ii), namely the wall is inside the active fluid.

The fitting analysis to the numerically evaluated efficiency factor tells us that for the three different cases, \eqref{efficiency factor kappav} can be well described by~\cite{Espinosa:2010hh}
\begin{align}
&\kappa_v(\alpha_\text{eff},v_w)= \frac{\alpha_\text{eff}}{\alpha}\nonumber\\
&\qquad
\times \begin{cases}
\frac{c_s^{11/5}\kappa_A\kappa_B}{(c_s^{11/5}-v_w^{11/5})\kappa_B+v_w c_s^{6/5}\kappa_A}\,, & \text{for  }v_w\lsim c_s\,,\\[1.5ex]
\kappa_B+(v_w-c_s)\delta\kappa + \frac{(v_w-c_s)^3}{(v_J-c_s)^3}[\kappa_C-\kappa_B-(v_J-c_s)\delta\kappa]\,, & \text{for }c_s \lsim v_w \lsim v_J\,,\\[1.5ex]
\frac{(v_J-1)^3v_J^{5/2}v_w^{-5/2}\kappa_C\kappa_D}{[(v_J-1)^3-(v_w-1)^3]v_J^{5/2}\kappa_C +(v_w-1)^3\kappa_D}\,,
& \text{for }v_J \lsim v_w\,,
\end{cases}
\label{general efficiency factor kappaV}
\end{align}
where the efficiency factors for different wall-velocity regions read
\begin{align}
\kappa_A&\simeq \frac{6.9\alpha_\text{eff}}{1.36-0.037\sqrt{\alpha_\text{eff}}+\alpha_\text{eff}}v_w^{6/5}\,, & &(v_w \ll c_s)\,, 
\notag \\[1ex]
\kappa_B&\simeq \frac{\alpha_\text{eff}^{2/5}}{0.017+(0.997+\alpha_\text{eff})^{2/5}}\,, & &(v_w = c_s)\,,
\notag \\[1ex]
\kappa_C&\simeq \frac{\sqrt{\alpha_\text{eff}}}{0.135+\sqrt{0.98+\alpha_\text{eff}}}\,,&  &(v_w=v_J)\,,
\notag \\[1ex]
\kappa_D&\simeq \frac{\alpha_\text{eff}}{0.73+0.083\sqrt{\alpha_\text{eff}}+\alpha_\text{eff}}\,,& &(v_w=1)\,.
\end{align}
In \eqref{general efficiency factor kappaV} for $c_s\lsim v_w \lsim v_J$, the efficiency factor $\kappa_v$ depends on $\delta \kappa$, which is the derivative of $\kappa_v$ with respect to $v_w$ at $v_w=c_s$, approximately given by
\begin{align}
\delta \kappa\simeq -0.9\log\left(\frac{\sqrt{\alpha_\text{eff}}}{1+\sqrt{\alpha_\text{eff}}}\right).
\end{align}
Then the efficiency factors $\kappa_\text{sw}$ and $\kappa_\text{turb}$ are defined by \eqref{efficiencies for SW and turb}.

In many previous works, the setup $\kappa_\text{sw}=\kappa_v$ and $\kappa_\text{turb}=\epsilon\kappa_v$ has been commonly employed, where $\epsilon$ is the fraction of turbulent bulk motion and is typically set to $\epsilon\simeq 0.05$ -- $0.1$, which stems from numerical simulations~\cite{Hindmarsh:2015qta}. The recent investigations \cite{Ellis:2018mja,Ellis:2019oqb,Ellis:2020awk}, however, have shown that when the period of the active production from sound waves is shorter than the Hubble time, the sound-wave GW amplitude is reduced and the GW from turbulence instead becomes stronger. This could be taken into account by introducing the parameter $H(T_p) \tau_\text{sw}$ as given in \eqref{efficiencies for SW and turb}. In this paper, we employ the recent treatment by following Ref.~\cite{Ellis:2019oqb,Ellis:2020awk}.

\section{Thin-wall approximation}
\label{App:ThinWall}
In this appendix, we discuss the thin-wall approximation that allows us to derive analytic approximations of $\alpha$ defined in \eqref{eq:alpha-theta}, the strength parameter of the first-order phase transition, and $\tilde\beta$ defined in \eqref{eq:LinearBeta3}, the inverse duration time (divided by the Hubble parameter), as functions of $\phi_c/T_c$. The thin-wall approximation corresponds to the limit $\varepsilon\to0$, where $\varepsilon(T_p)=V_\text{eff}(0,T_p)-V_\text{eff}(\langle\phi\rangle_{T_p},T_p)$ is the depth of the effective potential at the true vacuum at the percolation temperature as shown in \autoref{Thin-wall}.\footnote{The thin-wall approximation to the evaluation of the Euclidean action at zero temperature was discussed by Coleman~\cite{Coleman:1977py,Callan:1977pt,Coleman:1977th} and it was extended to the finite temperature case by Linde~\cite{Linde:1981zj}. The discussion in this appendix follows the latter.} This approximation holds in the weak supercooling regime where the percolation temperature is close to the critical temperature, $T_p\simeq T_c$. We normalize the effective potential at $\phi=0$ so that $V_\text{eff}(0,T)=0$ and thus $\varepsilon(T_p)=-V_\text{eff}(\langle\phi\rangle_{T_p},T_p)$.

\subsection{Latent heat and duration time in the thin-wall approximation}
The thin-wall approximation holds for weak supercooling, where the vacuum-energy density is negligible and \eqref{basic quantity: alpha: free energy} reduces to 
\begin{align}\label{eq:SuperCoolAlpha} 
\alpha\simeq   \frac{L_c}{4 \rho_{\textrm{rad}}}  = \frac{15}{2 \pi^2}\frac{L_c}{g_*(T_c)T_c^4}\,, 
\end{align}
with the latent heat $L_c$ at the critical temperature $T_c$,
\begin{align}\label{eq:LatentHeat}
L_c \equiv -  T\frac{\partial \Delta V_\text{eff}(\langle\phi\rangle_T,T)}{\partial T}\bigg|_{T=T_c}
= T\frac{\partial V_\text{eff}(\langle\phi\rangle_T,T)}{\partial T}\bigg|_{T=T_c}\, ,
\end{align}
where $\Delta V_\text{eff}(\langle\phi\rangle_T,T)=\varepsilon(T)=-V_\text{eff}(\langle\phi\rangle_{T},T)$. Next, we evaluate the duration time \eqref{eq:LinearBeta3} from the bounce solution to \eqref{bounce equation}. Within the thin-wall approximation, the bounce solution is approximately given by the case where the two vacua are degenerate.  
This is realized when the ``friction term'', the second term on the left-hand side in \eqref{bounce equation}, is negligible, to wit 
\begin{figure}
	\centering
	\includegraphics[width=.8\linewidth]{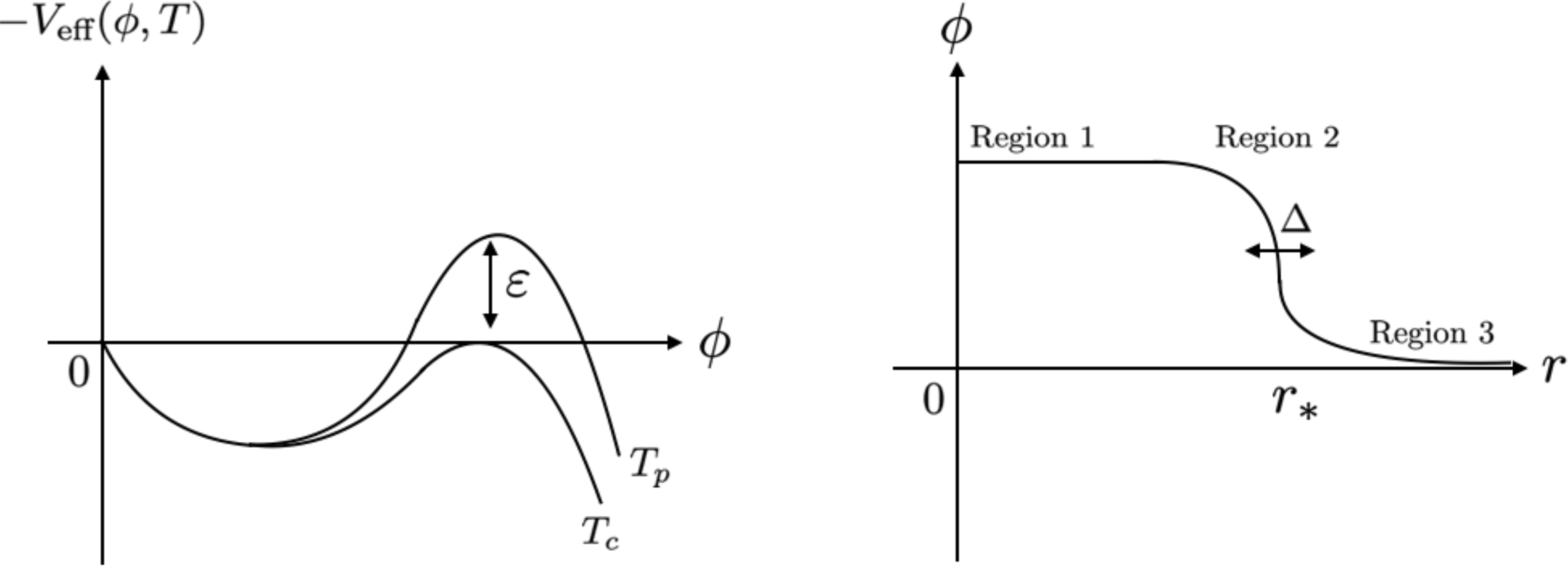}
	\caption{Schematic figures of the effective potential at $T_c$ and $T_p$, and the bounce solution to the equation of motion \eqref{EoM in the thin-wall approximation}.}
	\label{Thin-wall} 
\end{figure}
\begin{align}
\frac{{\mathrm d}^2\phi}{{\mathrm d} r^2} =\frac{\partial V_\text{eff}}{\partial \phi}\,.
\label{EoM in the thin-wall approximation}
\end{align}
This equation can be rewritten as
\begin{align}
\frac{{\mathrm d}\phi}{{\mathrm d} r} =-\sqrt{2V_\text{eff}(\phi,T)}\,.
\label{the equation of motion in the thin-wall approximation}
\end{align}
Its solution reads
\begin{align}
r=\int^{\phi_c}_{\phi} \! \frac{{\mathrm d} \varphi}{\sqrt{2V_\text{eff}(\varphi,T)}}\,.
\end{align}
For the evaluation of this integral we split the integration domain in three parts, see the right panel of \autoref{Thin-wall}, 
\begin{align}
\phi(r)=\begin{cases}
\phi_* & (0 < r < r_*-\Delta/2)~~~~\text{(Region 1)} \\[1ex]
\phi_w(r) & (r_*-\Delta/2 < r < r_*+\Delta/2)~~~~\text{(Region 2)} \\[1ex]
0& (r>r_*+\Delta/2)~~~~\text{(Region 3)}
\end{cases}.
\label{phir as function of r in the thin-wall approximation}
\end{align}
Here $\phi_*$ and $r_*$ are the field value and the radius of a bubble at a certain temperature $T$ around $T_c$, respectively, and $\Delta$ is the width of the bubble wall. Within the approximation \eqref{phir as function of r in the thin-wall approximation}, the spatial integral of the three-dimensional Euclidean action \eqref{three dimensional Euclidean action} reads, 
\begin{align}
S_3(T) &= 4\pi \int^\infty_0 {\mathrm d} r\,r^2\left[ \frac{1}{2}\left( \frac{{\mathrm d} \phi}{{\mathrm d} r} \right)^2 +V_\text{eff}(\phi,T)\right]  \nonumber\\
&=\underbrace{-\frac{4\pi}{3}r_*^3 \varepsilon(T)}_{\text{Region 1}} 
+ \underbrace{4\pi \int^{r_*+\Delta/2}_{r_*-\Delta/2}{\mathrm d} r\,r^2 \left[ \frac{1}{2}\left( \frac{{\mathrm d} \phi_w}{{\mathrm d} r} \right)^2 +V_\text{eff}(\phi_w,T)\right]}_{\text{Region 2}}  
+ \underbrace{0}_{\text{Region 3}}.
\end{align}
The region 2 is evaluated as
\begin{align}
(\text{Region 2})&= 4\pi r_*^2\int^{r_*+\Delta/2}_{r_*-\Delta/2}{\mathrm d} r \left[\frac{1}{2}\left( \frac{{\mathrm d} \phi_w}{{\mathrm d} r} \right)^2 +V_\text{eff}(\phi_w,T) \right] \nonumber\\[1ex]
&= 4\pi r_*^2\int^{\phi_*}_{0}{\mathrm d}\varphi \sqrt{2V_\text{eff}(\varphi,T)}
\equiv 4\pi r_*^2 S_1(T),
\label{S1 action in the thin-wall approximation}
\end{align}
where we used \eqref{the equation of motion in the thin-wall approximation} in the second equality. Here, $S_1$ corresponds to the bubble-wall tension $\sigma$ introduced in \eqref{Eq: Lagrangian of bubble radius}. From the stationary condition 
\begin{align}
\frac{{\mathrm d} S_3(T)}{{\mathrm d} r_*}=-4\pi r_*^2 \varepsilon(T) +8\pi r_* S_1(T) =0\,,
\end{align}
one finds
\begin{align}
r_*&= \frac{2S_1(T)}{\varepsilon(T)}\,,
&
S_3(T)&=\frac{16\pi }{3}\frac{S_1(T)^3}{\varepsilon(T)^2}\,.
\label{three dimensional Euclidean action in the thin-wall approximation}
\end{align}
Note that from these relations, one obtains
\begin{align}
r_* = \left[ \frac{3S_3}{2\pi \varepsilon}\right]^{\frac{1}{3}}.
\end{align}
This result is compatible with the critical radius $r_c=2\sigma/p$ in \autoref{App: definitions of efficiencies} by inserting $p=\varepsilon$ and $\sigma=S_1$. For weak supercooling the percolation temperature, $T_p< T_c$, is close to the critical temperature. For temperatures $T_p\leq T\leq T_c$, this entails a small reduced temperature $(T-T_c)/T_c \ll 1$. This allows us to use a linear approximation for the expansion of the effective potential about $T_c$ or rather in powers of the reduced temperature. The linear expansion coefficient in an expansion in powers of the reduced temperature is the latent heat,  \eqref{eq:LatentHeat}, and we arrive at 
\begin{align}\label{eq:ExpandEffPot}
V_\text{eff}(\phi,T) = V_\text{eff}(\phi_c,T_c) +T \frac{ \partial V_\text{eff}}{\partial T}\bigg|_{T=T_c}\frac{T-T_c}{T_c} +\mathcal{O}\!\left(\left[\frac{T-T_c}{T_c}\right]^2\right)\simeq   L_c\, \frac{T-T_c}{T_c} .
\end{align}
In \eqref{eq:ExpandEffPot} we have used that $V_\text{eff}(\phi_c,T_c)=0$ at $T_c$. For $T=T_p$ this leads us to 
\begin{align}
&V_\text{eff}(\phi_*,T_p)=\varepsilon(T_p)\simeq -L_c \delta_c(T_p),
\quad \text{with}\quad \delta_c(T) = \frac{T_c-T}{T_c}\,.
\label{eq: expansion of potential around Tc}
\end{align}
Their insertion into the three dimensional Euclidean action \eqref{three dimensional Euclidean action in the thin-wall approximation} yields, at a certain $T$ around $T_c$,
\begin{align}
\frac{S_3(T)}{T} &\simeq \frac{16\pi}{3}\frac{S_1(T_c)^3}{T_cL_c^2 }\frac{\delta_c(T)^{-2}}{1-\delta_c(T)}\left( 1-3\delta_c(T)\frac{T_c}{S_1(T_c)}\frac{\partial S_1(T)}{\partial T}\bigg|_{T=T_c} \right)\nonumber\\[1ex]
&\simeq \frac{16\pi}{3}\frac{S_1(T_c)^3}{T_cL_c^2 }\delta_c(T)^{-2} ,
\label{eq: three dimensional Euclidean action in terms of Tc}
\end{align}
from which one finds
\begin{align}
\delta_c(T) =\left(\frac{16\pi}{3}\frac{S_1(T_c)^3}{T_c L_c^2} \right)^{1/2}\left(\frac{S_3(T)}{T}\right)^{-1/2}.
\end{align}
The duration time \eqref{eq:LinearBeta3} at the percolation temperature reads
\begin{align}
\tilde\beta&
=T\frac{\mathrm d}{\mathrm dT}\frac{S_3(T)}{T}\bigg|_{T=T_p} 
\simeq \frac{16\pi}{3}\frac{S_1(T_c)^3}{T_c L_c^2 }\frac{\mathrm d }{\mathrm d T}\delta_c(T)^{-2}\bigg|_{T=T_p}\nonumber\\[1ex]
&= \frac{2}{\delta_c(T_p)}\frac{S_3(T_p)}{T_p} 
=\left(\frac{3}{4\pi} \frac{T_cL_c^2}{S_1(T_c)^3}  \right)^{1/2}\left(\frac{S_3(T_p)}{T_p}\right)^{3/2}.
\label{eq: beta tilde in the thin-wall approximation}
\end{align}

\subsection{A simple model case: $\phi^4$ model}
\label{app:ThinWall4}
We consider the effective potential
\begin{align}
V_\text{eff}(\phi,T)=A(T^2- T_0^2)\phi^2 -BT \phi^3 +\frac{\lambda_T}{4}\phi^4,
\label{eq: starting phi4 potential}
\end{align}
where $A$, $B$, and $\lambda_T$ are positive constant and $T_0$ is the temperature at which the symmetric phase $\phi=0$ becomes metastable.
At $T_c$ or for $\varepsilon\to0$, the potential values at $\phi=0$ and $\phi=\langle\phi\rangle_T$ take the same value $V_\text{eff}=0$, so that in such a case one can parametrize the effective potential as
\begin{align}
V_\text{eff}(\phi,T_c)=\frac{\lambda_T}{4}\phi^2(\phi-\phi_c)^2.
\label{phi4 effective potential at Tc}
\end{align}
Comparing between \eqref{eq: starting phi4 potential} and \eqref{phi4 effective potential at Tc} at $T_c$, we obtain the relations
\begin{align}
&A(T_c^2-T_0^2)=\frac{\lambda_T}{4}\phi_c^2,&
&BT_c=\frac{\lambda_T}{2}\phi_c,&
&T_c^2\left( 1-\frac{B^2}{A\lambda_T} \right) =T_0^2.
\label{eq: relations between couplings and temperatures}
\end{align}
 Here, $T_0$ has to be positive, which implies that $A>B^2/\lambda_T$. For the effective potential \eqref{phi4 effective potential at Tc}, the bounce solution to the equation of motion \eqref{EoM in the thin-wall approximation} is found to be
\begin{align}
\phi(r)=\frac{\phi_c}{2}\left[ 1 -\tanh \left( \frac{r-r_*}{\Delta}\right)  \right],
\label{eq: bounce solution in the thin-wall approximation}
\end{align}
with $r_*$ the bubble wall radius and $\Delta=(2/\phi_c)\sqrt{2/\lambda_T}$ the bubble wall width. Inserting this bounce solution into \eqref{S1 action in the thin-wall approximation}, one obtains 
\begin{align}
S_1(T_p)&\simeq \int^\infty_0 {\mathrm d} r \left[\frac{1}{2}\left( \frac{{\mathrm d} \phi}{{\mathrm d} r}\right)^2 +\frac{\lambda_T}{4}\phi^2(\phi-\phi_c)^2 \right] \nonumber\\[1ex]
&= \frac{\lambda_T\phi_c^4 \Delta}{32} \int^\infty_{-r_*/\Delta} \frac{{\mathrm d} x}{\cosh^4 x} 
\simeq \frac{\sqrt{2}}{12}\lambda_T^{1/2} \phi_c^3,
\end{align}
where we have employed the thin-wall approximation, namely $-r_*/\Delta\to -\infty$ in the last equality and have used
\begin{align}
\int^\infty_{-\infty} \frac{{\mathrm d} x}{\cosh^4 x} =\frac{4}{3}.
\end{align}
Note that instead of the use of the bounce solution \eqref{eq: bounce solution in the thin-wall approximation}, we can obtain the same result by evaluating directly \eqref{S1 action in the thin-wall approximation} with the potential \eqref{phi4 effective potential at Tc}, 
\begin{align}
S_1(T_p)&=\int^{\phi_c}_{0}{\mathrm d}\varphi \sqrt{2V_\text{eff}(\varphi,T_p)} =  \frac{\sqrt{2}}{12}\lambda_T^{1/2} \phi_c^3.
\end{align}
Thus, the latent heat $L_c$, $\alpha$ and $\tilde\beta$ in this simple model are evaluated at $T_p\simeq T_c$, respectively as
\begin{align}
L_c&=T \frac{\partial V_\text{eff}}{\partial T}\bigg|_{T=T_c}
=2\left(A-\frac{B^2}{\lambda_T}\right) T_c^4 \left(\frac{\phi_c}{T_c}\right)^2,
\notag \\[1ex]
\alpha&\simeq 
\frac{15}{\pi^2 g_* (T_c)}\left(A-\frac{B^2}{\lambda_T}\right)  \left(\frac{\phi_c}{T_c}\right)^2
\equiv C_\alpha \left(\frac{\phi_c}{T_c}\right)^2,
\notag \\[1ex]
\tilde\beta&\simeq
\frac{36\cdot 2^{1/4}}{\pi^{1/2}\lambda^{3/4}_T} \left(A-\frac{B^2}{\lambda_T}\right) \left( \frac{S_3(T_p)}{T_p}\right)^\frac{3}{2} \left( \frac{\phi_c}{T_c}\right)^{-5/2}
\equiv C_{\tilde\beta} \left(\frac{\phi_c}{T_c}\right)^{-5/2},
\end{align} 
where we have used \eqref{eq: relations between couplings and temperatures}. The positivity of the latent heat implies that $A>B^2/\lambda_T$, which is the same condition as the positivity of $T_0$. As discussed in \eqref{criterion for nucleation temperature in terms of action}, the factor $S_3(T_p)/T_p$ takes a constant value between $140$--$150$ when the weak supercooling occurs at $T_p\approx T_n$. We see now that $\alpha$ behaves like $(\phi_c/T_c)^2$, while $\tilde\beta$ behaves like $(\phi_c/T_c)^{-5/2}$. From this fact, $\tilde\beta$ behaves as a function of $\alpha$ such that
\begin{align}
\tilde\beta=  C \alpha^{-5/4},
\end{align}
where $C=C_{\tilde\beta}/C_\alpha^{-5/4}$. 

\begin{landscape}
\section{Result tables}
\label{app:res-tables}

\begin{table}[htbp]
\begin{tabular}{c|c|c|c|c|c|c|c|c|c|c|c|c}
$\lambda_6$ & $\phi_c/T_c$ & $\alpha$ & $\tilde \beta$ & $R\cdot$GeV & $H(T_p)/$GeV & $T_c/$GeV & $T_n/$GeV & $T_p/$GeV & $T_\text{reh}/$GeV & SNR& $\lambda_{H^3}/\lambda_{H^3,0}$ & $\lambda_{H^4}/\lambda_{H^4,0}$\\
\hline
\hline
 58 & 1.02 & 0.00274 & 46400 & $3.40\cdot 10^{9}$ & $1.86 \cdot 10^{-14}$ & 116 & 115 & 115 & 115 & $1.97\cdot 10^{-20}$ & 1.67 & 5.05  \\
 \hline
 64 & 1.26 & 0.00436 & 19200 & $8.91\cdot 10^{9}$ & $1.71 \cdot 10^{-14}$ & 112 & 111 & 110 & 111 & $6.37\cdot 10^{-15}$  & 1.73 & 5.43 \\
 \hline
 72 & 1.50 & 0.00655 & 8950 & $2.12\cdot 10^{10}$ & $1.54 \cdot 10^{-14}$ & 107 & 105 & 105 & 105 & $2.33\cdot 10^{-12}$  & 1.81 & 5.88 \\
 \hline
 82 & 1.83 & 0.0110 & 4260 & $5.63\cdot 10^{10}$ & $1.22 \cdot 10^{-14}$ & 97.8 & 93.6 & 93.2 & 93.5 & $4.06 \cdot 10^{-10}$ & 1.90 & 6.42  \\
 \hline
 93 & 2.03 & 0.0156 & 2520 & $1.05\cdot 10^{11}$ & $1.11 \cdot 10^{-14}$ & 96.1 & 89.5 & 88.9 & 89.3 & $9.84\cdot 10^{-8}$  & 2.01 & 7.02 \\
 \hline
 101 & 2.20 & 0.0215 & 1670 & $1.82\cdot 10^{11}$ & $9.61 \cdot 10^{-15}$ & 92.4 & 83.4 & 82.6 & 83.0 & $3.66\cdot 10^{-6}$ & 2.08 & 7.41 \\
 \hline
 104 & 2.27 & 0.0251 & 1400 & $2.35\cdot 10^{11}$ & $8.92 \cdot 10^{-15}$ & 90.7 & 80.4 & 79.5 & 80.0 & 0.0000714 & 2.11 & 7.58 \\
 \hline
 112 & 2.42 & 0.0365 & 917 & $4.27\cdot 10^{11}$ & $7.48 \cdot 10^{-15}$ & 87.5 & 73.8 & 72.6 & 73.3 & 0.000267 & 2.17 & 7.93 \\
 \hline
 124 & 2.68 & 0.0846 & 394 & $1.49\cdot 10^{12}$ & $4.98 \cdot 10^{-15}$ & 82.1 & 60.6 & 58.7 & 59.9 & 0.00541 & 2.27 & 8.49 \\
 \hline
 130 & 2.84 & 0.232 & 127 & $7.24\cdot 10^{12}$ & $3.18 \cdot 10^{-15}$ & 78.9 & 48.6 & 45.6 & 48.0 & 1.40 & 2.32 & 8.79 \\
 \hline
 132 & 2.87 & 0.344 & 67.0 & $1.61\cdot 10^{13}$ & $2.72 \cdot 10^{-15}$ & 78.4 & 44.5 & 41.2 & 44.3 & 126 & 2.33 & 8.85 \\
 \hline
 132.2 & 2.89 & 0.518 & 21.7 & $5.69\cdot 10^{13}$ & $2.37 \cdot 10^{-15}$ & 78.0 & 41.8 & 37.3 & 41.4 & 379 & 2.33 & 8.88 \\
 \hline\hline
 132.6 & 2.90 & 0.817 & 19.3 & $1.22\cdot 10^{14}$ & $2.07 \cdot 10^{-15}$ & 77.8 & 39.1 & 33.3 & 38.6 & 675 & 2.34 & 8.90 \\
 \hline
 132.9 & 2.90 & 0.966 & 19.3 & $1.74\cdot 10^{14}$ & $1.99 \cdot 10^{-15}$ & 77.8 & 38.2 & 32.0 & 37.9 & 811 & 2.34 & 8.90 \\
\end{tabular}
\caption{The results for the $\phi^6$ modification of the Higgs potential are summarized. The double line indicates where the strong supercooling regime, i.e., the regime where a minimization temperature exists as described in \autoref{sec:strong-supercooling}.}
\label{tab:results-poly}
\end{table}

\begin{table}[htbp]
	\begin{tabular}{c|c|c|c|c|c|c|c|c|c|c|c|c}
	$\lambda_{\text{ln}}$& $\phi_c/T_c$ & $\alpha$ & $\tilde \beta$ & $R\cdot$GeV & $H(T_p)/$GeV & $T_c/$GeV & $T_n/$GeV & $T_p/$GeV & $T_\text{reh}/$GeV & SNR & $\lambda_{H^3}/\lambda_{H^3,0}$ & $\lambda_{H^4}/\lambda_{H^4,0}$\\
	\hline
	\hline
	0.36 & 1.03 & 0.00286 & 37400 & $4.13\cdot 10^9$ & $1.90\cdot 10^{-14}$ & 117 & 116 & 116 & 116 & $2.05\cdot 10^{-14}$  & 1.58 & 4.19\\
	\hline
	0.4& 1.27 & 0.00450 & 16400 & $1.03\cdot 10^{10}$ & $1.74\cdot 10^{-14}$ & 113 & 111 & 111 & 111 & $5.51\cdot 10^{-12}$  & 1.65 & 4.53 \\
	\hline
	0.51& 1.76 & 0.0102 & 4290 & $5.08\cdot 10^{10}$ & $1.35\cdot 10^{-14}$  & 103 & 98.3 & 97.8 & 98.1 & $6.63\cdot 10^{-8}$ & 1.82 & 5.43\\
	\hline
	0.58 & 2.03 & 0.0164 & 2210 & $1.20\cdot 10^{11} $ & $1.10\cdot 10^{-14}$ & 96.5 & 89.2 & 88.5 & 88.8 & $7.75\cdot 10^{-6}$ & 1.92 & 5.99\\
	\hline
	0.64& 2.25 & 0.0259 & 1280 & $2.58\cdot 10^{11}$ & $8.90\cdot 10^{-15}$ & 91.6 & 80.3 & 79.4 & 79.9 &  $0.000429$ & 2.01 & 6.44\\
	\hline
	0.65& 2.27 & 0.0273 & 1200 & $2.80\cdot 10^{11}$ &  $8.70\cdot 10^{-15}$ & 91.0 & 79.5 & 78.5 & 79.0 & $0.000661$  & 2.02 & 6.51\\
	\hline
	0.66 & 2.32 & 0.0304 & 1070 &  $ 3.33\cdot 10^{11}$  &  $8.26\cdot 10^{-15}$  & 89.9 & 77.5 & 76.4 & 76.4 & 0.00159  & 2.04 & 6.60 \\
	\hline
	0.71& 2.51 & 0.0503 & 628 & $7.20\cdot 10^{11}$  &   $6.47\cdot 10^{-15}$ & 85.9 & 68.9 & 67.4 & 68.2 & 0.0640  & 2.11 & 6.97\\
	\hline
	0.735 & 2.59 & 0.0680 & 464 & $1.13\cdot 10^{12}$  &  $5.61\cdot 10^{-15}$ & 84.2 & 64.3 & 62.5 & 63.6 & 0.448 & 2.14 & 7.13\\
	\hline
	0.76 & 2.67 & 0.100 & 313 &  $2.00\cdot 10^{12}$ & $4.69\cdot 10^{-15}$ & 82.5 & 58.9 & 56.8 & 58.2 & 4.04 & 2.17 & 7.30\\
	\hline
	0.765 & 2.70 & 0.116 & 269 &  $2.49\cdot 10^{12}$ & $4.38\cdot 10^{-15}$ & 81.9 & 57.0 & 54.7 & 56.2 & 8.42 & 2.18 & 7.34\\
	\hline
	0.7825 & 2.77 & 0.196 & 147 & $5.72\cdot 10^{12}$  &  $3.48\cdot 10^{-15}$  & 80.4 & 50.8 & 48.0 & 50.2 & 75.8 & 2.20 & 7.47\\
	\hline\hline
	0.8 & 2.85 & 1.53 & 19.8 & $3.48\cdot 10^{14}$ &  $1.83\cdot 10^{-15}$  & 78.8 & 36.3 & 28.8 & 36.4 & 383  & 2.23 & 7.61\\
\end{tabular}
\caption{The results for the $\phi^4\log(\phi^2)$ modification of the Higgs potential are summarized. The double line indicates where the strong supercooling regime, i.e., the regime where a minimization temperature exists as described in \autoref{sec:strong-supercooling}.}
\label{tab:results-log}
\end{table}

\begin{table}
\begin{tabular}{c|c|c|c|c|c|c|c|c|c|c|c|c}
	$\lambda_{\text{exp}}/10^{10}$ &$\phi_c/T_c$ & $\alpha$ & $\tilde \beta$ & $R\cdot$GeV & $H(T_p)/$GeV & $T_c/$GeV & $T_n/$GeV & $T_p/$GeV & $T_\text{reh}/$GeV & \text{SNR} & $\lambda_{H^3}/\lambda_{H^3,0}$ & $\lambda_{H^4}/\lambda_{H^4,0}$\\
	\hline
	\hline
	$1.3$& 1.05 & 0.00292 & 65200 & $2.67\cdot 10^9$ & $1.68\cdot10^{-14}$ & 110 & 110 & 109 & 110 & $1.62\cdot 10^{-15} $ & 2.00 & 8.72  \\
	\hline
	$1.5$& 1.17 & 0.00362 & 44500 & $4.00\cdot 10^9$ & $1.65\cdot 10^{-14}$ & 109 & 108 & 108 & 108 & $2.14\cdot 10^{-14}$ & 2.02 & 8.86 \\
	\hline
	$1.6$& 1.24 & 0.00406 & 35400 & $5.12\cdot 10^9$ & $1.62\cdot 10^{-14}$ & 108 & 107 & 107 & 107 & $9.80\cdot 10^{-14}$ & 2.04 & 8.96 \\
	\hline
	$2.3$& 1.51 & 0.00630 & 14900 & $1.33\cdot 10^{10}$ & $1.48\cdot 10^{-14}$ & 104 & 103 & 103 & 103 & $3.11\cdot 10^{-11}$ & 2.11 & 9.44 \\
	\hline
	3.5& 1.76 & 0.00889 & 9470 & $2.34\cdot 10^{10}$ & $1.32\cdot10^{-14}$ & 99.9 & 97.3 & 97.0 & 97.3 & $9.46\cdot 10^{-10}$ & 2.19 & 9.97 \\
	\hline
	5.4 & 1.99 & 0.0129 & 4180 & $6.03\cdot 10^{10}$ & $1.16 \cdot 10^{-14}$ & 95.7 & 91.3 & 90.9 & 91.2 & $1.83\cdot 10^{-7}$ & 2.27 & 10.5 \\
	\hline
	8.2& 2.21 & 0.0184 & 2490 & $1.18\cdot 10^{11}$ & $9.97\cdot 10^{-15}$ & 91.4 & 84.7 & 84.1 & 84.5 & $7.28\cdot 10^{-6}$ & 2.36 & 11.1 \\
	\hline
	14.5& 2.51 & 0.0330 & 1230 & $3.15\cdot 10^{11}$ & $7.58\cdot 10 ^{-15}$ & 85.6 & 74.1 & 73.2 & 73.8 & 0.00121 & 2.48 & 11.9 \\
	\hline
	23.8& 2.78 & 0.0703 & 573 & $9.74\cdot 10^{11}$ & $5.25\cdot10^{-15}$ & 80.3 & 61.9 & 60.5 & 61.5 & 0.259 & 2.58 & 12.6\\
	\hline
	33 & 2.95 & 0.161 & 258 & $3.16\cdot 10^{12}$ & $3.59\cdot 10^{-15}$ & 77.1 & 51.3 & 49.1 & 51.0 & 21.0 & 2.65 & 13.0 \\
	\hline
	33.8& 2.97 & 0.191 & 213 & $4.14\cdot 10^{12}$ & $3.33\cdot 10^{-15}$ & 76.1 & 49.3 & 47.0 & 49.1 & 44.2 & 2.65 & 13.0\\
	\hline
	35.4& 3.00 & 0.248 & 161 & $6.09\cdot 10^{12}$ & $2.99\cdot 10^{-15}$ & 75.2 & 46.6 & 44.0 & 46.5 & 112 & 2.66 & 13.1\\
	\hline
	39& 3.05 & 0.553 & 45.8 & $2.86\cdot 10^{13}$ & $2.24\cdot 10^{-15}$ & 74.8 & 39.8 & 36.1 & 40.3 & 846 & 2.68 & 13.2\\
	\hline\hline
	40.2& 3.07 & 1.34 & 18.5 & $2.25\cdot 10^{14}$ & $1.77\cdot 10^{-15}$ & 76.6 & 34.6 & 28.9 & 35.8 & 787 & 2.69 & 13.2\\
\end{tabular}
\caption{The results for the $\phi^4 \exp(-1/\phi^2)$ modification of the Higgs potential are summarized. The double line indicates where the strong supercooling regime, i.e., the regime where a minimization temperature exists as described in \autoref{sec:strong-supercooling}.}
\label{tab:results-exp}
\end{table}

\end{landscape}

\bibliographystyle{JHEP} 
\bibliography{refs}

\end{document}